\documentclass[a4paper,reqno,12pt,draft]{article}

\begin{document}

\begin{center}

{\LARGE {Enhanced Coset Symmetries and Higher Derivative Corrections}}\\

\bigskip

\Large
{Neil Lambert}\footnote{neil.lambert@kcl.ac.uk}
and {Peter West}\footnote{pwest@mth.kcl.ac.uk}\\

Dept. of Mathematics\\
King's College\\
The Strand\\
London\\
WC2R 2LS, UK\\

\end{center}

\bigskip
\begin{center}
{\bf { Abstract}}
\end{center}

After dimensional reduction to three dimensions, the lowest order
effective actions for pure gravity, M-theory and the Bosonic string
admit an enhanced symmetry group. In this paper we initiate study of
how this enhancement is affected by the inclusion of higher
derivative terms. In particular we show that the coefficients of the
scalar fields associated to the Cartan subalgebra are given by
weights of the enhanced symmetry group.

\newpage

\section{ Introduction}

Classical supergravity actions have played a central role in
theoretical high energy physics and string theory in particular. One
unexpected development was the realization that these theories lead
to enhanced symmetries when dimensionally reduced on a torus. This
phenomenon was first observed in the dimensional reduction of
four-dimensional Einstein gravity to three dimensions where the
resulting theory was found to have an $SL(2)$ symmetry
\cite{Ehlers}. More generally, gravity reduced to three dimensions
on an $n$-torus leads to a theory with an $SL(n+1)$ symmetry. The
$SL(n)$ part of this symmetry is just the diffeomorphisms of the
higher dimensional theory that are unbroken by the torus. The
enhancement to $SL(n+1)$ only arises when one dualizes the rank two
field strength of the graviphotons (the components of the metric
that have an internal and space-time index) to obtain scalars in
three dimensions. These scalars, together with those which arise
directly from the metric, form  a non-linear realization of  $SL
(n+1)$ with  local subalgebra $SO(n+1)$.

The dimensional reduction of eleven dimensional supergravity, or the
type II supergravities, on a torus to four and three dimensions
leads to a $E_7$ \cite{Cremmer:1978ds} and $E_8$
\cite{Marcus:1983hb} symmetry respectively. Furthermore, the
effective theory  associated with the Bosonic string, which consists
of gravity, a scalar and a rank two gauge field, when dimensional
reduced on a $n$ torus to three dimensions leads to an $O(n+1,n+1)$
symmetry. For other examples see \cite{Breitenlohner:1987dg}.

In this paper we will only consider the Bosonic sector of the
effective action. However it would be most interesting to extend our
analysis to include Fermions. The dimensional reduction to three
dimensions is special in that it is the first dimension in which all
fields of the original theory become scalars after the appropriate
dualizations. Thus the resulting theory only consists of scalars
and, for all the theories mentioned above, is a non-linear
realization of a group $G$ with local sub-algebra $H$. Since the
action is only second order in space-time derivatives it is given by
\begin{equation} S = \int d^3 x\sqrt{-g}\left(R - {\rm
Tr}(g^{-1}D_\mu g g^{-1}D^\mu g) \right)
\label{GHaction}
\end{equation}
where $g \in G$ and $D_\mu$ is the $H$-covariant derivative. This
action has the manifest symmetry $g\to g_0 g h$ where $g_0\in G$ is
a constant transformation, (i.e. rigid transformation) and $h\in H $
is local. In fact in all the above cases the local subalgebra $H$ is
the just the subalgebra of $G$ that is invariant under the Cartan
involution. A generic theory, when dimensionally reduced, will not
possess an enhanced symmetry algebra and so will not be expressible
as a non-linear realization of the above form. Indeed, it requires a
very precise field content with corresponding couplings to find such
a symmetry \cite{Cremmer:1999du,Lambert:2001gk}.

To show that the  theories mentioned above,   when dimensionally
reduced to three dimensions, have the action of equation
(\ref{GHaction}) is in general quite technically complicated.
However, the above action has a particularly characteristic
expression. Taking account of the local transformations the group
element may be chosen to be
\begin{equation}
g = e^{\sum_{\underline \alpha>0}
\chi_{\underline\alpha}E_{\underline \alpha}} \
e^{-\frac{1}{\sqrt{2}}\underline\phi \cdot \underline H}
\end{equation}
where $ \underline H$ and $E_{\underline \alpha}$ are the Cartan
subalgebra and positive root generators of $G$ respectively. One
then finds that the coset part of the action has the form
\begin{equation}
{\rm Tr}(g^{-1}D_\mu g g^{-1}D^\mu g) = \frac{1}{2}\partial_\mu
\underline\phi \cdot\partial^\mu \underline\phi
+\frac{1}{2}\sum_{\underline \alpha>0}
e^{\sqrt{2}\underline\phi\cdot \underline\alpha}
\partial_\mu\chi_{\underline\alpha}\partial^\mu\chi_{\underline\alpha}+\ldots
\end{equation}
where the ellipsis denotes higher order terms in
$\chi_{\underline\alpha}$. Note that the roots of the algebra $G$
show up as vectors that occur in the exponentials that multiply the
kinetic fields $\chi_{\underline\alpha}$. These fields arise from
the off-diagonal metric components and the full effective action can
be complicated to evaluate exactly. On the other hand  the fields
$\underline \phi$ arise from the diagonal components of the metric
(and dilaton if present) and their role in the dimensional reduction
procedure is relatively easy to follow. As such it is
straightforward  to find out in which theories the roots of an
algebra arise and in addition what algebra they belong to.

The IIA \cite{Campbell:1984zc,Giani:1984wc,Huq:1983im} and IIB
\cite{Schwarz:1983wa,Howe:1983sr,Schwarz:1983qr} supergravity
theories  are the complete low energy effective actions for the IIA
and IIB superstring theories. As a result,  much of our
understanding of non-perturbative effects of string theory has been
derived from these theories. These theories are now believed to be
different perturbative descriptions of M-theory and their
dimensional reduction can be obtained considering the reduction of
eleven-dimensional supergravity \cite{CJS}.

Part of the symmetries that occur in the dimensional reduction are
T-dualities \cite{Kikkawa,Buscher,RV}, which have been shown to be a
symmetry of string theory at all orders of perturbation theory.
However, all the symmetries found in the dimensional reduction of
the supergravity theories, or more precisely a discrete subgroup
thereof, have been conjectured to be symmetries of string theory and
called U-duality \cite{HT}. In fact U-dualities can generated by
T-duality together with a discrete version of the $SL(2)$ symmetry
of IIB supergravity \cite{Schwarz:1983wa}.

It has been conjectured that M-theory possess an infinite
dimensional Kac-Moody symmetry,  traces of which can be found in the
type II supergravity theories \cite{West:2001as}. In particular, one
should consider the non-linear realization of $E_{11}$ with  the
local subalgebra being the Cartan involution invariant subalgebra.
From this viewpoint, the symmetries that arise upon  dimensional
reduction are not  accidents of dimensional reduction, as is
commonly believed, but are remnants of  the symmetries that occur in
the $E_{11}$ non-linearly realized theory before dimensional
reduction. A related approach \cite{Damour:2002cu}, which shares
some similar ideas, is based on $E_{10}$, a subalgebra of $E_{11}$.
The difference in approach can be traced to the discovery of
"cosmological billiards", or BKL behavior, that occurs when
eleven-dimensional supergravity is considered in the region of a
space-like singularity \cite{Damour:2000hv}. In a recent paper
\cite{DN} higher derivative terms were also considered within the
context of the BKL limit and it was found that the higher derivative
terms thought to occur in M-theory lead to the appearance of roots
of $E_{10}$ in the resulting cosmological billiards. However, it has
been suggested that these should be thought of as weights and
furthermore  weights should also arise when considering the BKL
limit of higher derivative terms in pure gravity \cite{Marc}.

Much of the discussion of these symmetries has been in the context
of the lowest order, two derivative,  effective action. An exception
is \cite{Meissner} which considers how T-duality is altered by the
leading order higher derivative corrections, although it does not
consider the enhancement that occurs in three dimensions. Indeed to
the best of our knowledge there has been little  or no discussion as
to whether or not the enhanced symmetries that occur in the
dimensional reduction of the low energy effective actions extend to
the higher derivative corrections. Therefore in this paper we begin
a systematic study how the enhanced symmetry is affected by higher
derivative terms.

In general higher derivative corrections are  very complicated and
only some specific terms are known \cite{GreenSchwarz}-\cite{Howe},
although the complete next-to-leading order result has recently been
reported \cite{Policastro:2006vt}. Hence the subject is not
sufficiently advanced at this stage to determine whether or not the
higher derivative terms, once dimensionally reduced to three
dimensions, possess an enhanced non-linearly realized symmetry.
However, for the generic higher derivative terms it is possible to
derive the dependence of the diagonal components of the metric (and
dilaton if present) in the dimensional reduction to three dimensions
and then read off their coefficients. These coefficients form a
constant vector. As explained above for the low energy effective
action at lowest order in derivatives, one finds the roots of the
enhanced symmetry algebra.

In this paper we will show that for gravity, M-theory and the
Bosonic string one finds weights of the enhanced symmetry algebra,
but only  for certain  classes of higher derivative terms. For
higher derivative terms outside these classes the vectors that arise
have no apparent interpretation in terms of objects associated with
the algebra. However the class of higher derivative terms for which
weights arise are just those that are expected on the basis of
string theory arguments. Demanding the appearance  of weights  can
therefore  be used to predict the correct class of higher derivative
terms. The appearance of weights provides strong evidence for some
enhanced symmetry structure in the higher derivative terms. On the
other hand, the appearance of weights, as opposed to roots, is
intriguing as the Cartan forms that are usually used to construct
actions for non-linearly realized theories only contain the roots.

The rest of this paper is organized as follows. In section two we
consider the case of pure gravity and recall the derivation of the
$SL(n+1)$ symmetry for the reduction of gravity on an $n$ torus to
three dimensions. We then  introduce in detail a new technique for
deriving the dependence of the higher order effective action on the
diagonal components of the metric, ({\it i.e.} the fields associated
with the Cartan subalgebra). We then show that higher derivative
terms lead to weights of the extended $SL(n+1)$ symmetry appearing
as the coefficients of the scalar fields which form the diagonal
components of the internal metric. In the section three we repeat
this analysis for M-theory. Here we find that the weights of $E_8$
appear only if the higher derivative terms have  the form $(\hat
R)^{3k+1}$, for $k=0,1,2,...$, as is expected from quantum string
calculations. In section four we consider the Bosonic string and our
analysis shows that the weights of $O(n+1,n+1)$ appear only if the
terms have a particular power of the dilaton which in turn
corresponds to a particular genus contribution of a  perturbative
string calculation. In section four we discuss the effect of having
more than one derivative acting on each field. In section five we give a conclusion.
The paper contains also three appendices which
discuss technical details that are used through the main text. In
particular the first gives details of the dimensional reduction, the
second recalls some facts about non-linear realizations and the
third some of the required theory of group representations.

\section{Pure Gravity}

In this section we will discuss pure gravity,
compactified on and $n$-torus
from  $D$-dimensions to 3-dimensions. Our main motivation is
to develop the techniques that we will use later.
We start with the dimensional reduction of the pure gravitational action
\begin{equation}
S  = \int d^Dx \sqrt{-\hat g}\hat R
\end{equation}
where we use a hat to denote $D$-dimensional quantities.
We write the $D$-dimensional metric as
\begin{equation}
d\hat s^2 = e^{2\alpha \rho}ds^2 + e^{2\beta\rho}G_{ij}
(dx^i + A^i_\mu dx^\mu)
(dx^j + A^j_\mu dx^\mu)
\end{equation}
and we refer the reader to Appendix A for details of our conventions and
various formulae.
Following these calculations we find that the Einstein-Hilbert
action, dimensionally reduced to three dimensions in Einstein
frame, is
\begin{equation}
S = \int d^3x \sqrt{-g}\Big(
R - \frac{1}{4}e^{-2\frac{n+1}{n}\alpha\rho} G_{ij}F^i_{{\mu\nu}} F^{{j\mu\nu}}
- {\rm Tr}(S_\mu S^\mu)-\frac{1}{2}\partial_\mu\rho\partial^\mu\rho\Big)
\end{equation}
Here $S_{\mu\ \overline i}^{\ \overline j}$ is the symmetric part of
$e_{\overline i}^{\ k}\partial_\mu e_k^{\ \overline j}$ where $
e_i^{\ \overline j}$ is the vielbein for the internal metric
$G_{ij}$ and an over-line denotes a tangent frame index.

Here we see that  the action has a manifest $SL(n)/SO(n)$ symmetry
that acts on the internal vielbein and rotates the graviphotons (we
refer the reader to Appendix B for a review of $G/H$ coset
Lagrangians). In particular the action of  $SL(n)$ corresponds to
diffeomorphisms that preserve the torus whereas the $SO(n)$
generates local tangent frame rotations.

But wait there's more!
To see this we can add
the Lagrange multiplier term associated to the Bianchi identity of
the graviphotons
\begin{eqnarray}
S_{LM} &=& \frac{1}{2}\int d^3x\epsilon^{\mu\nu\lambda}\chi_i\partial_\mu F^i_{\nu\lambda}\nonumber\\
 &=&  -\frac{1}{2}\int d^3x\epsilon^{\mu\nu\lambda}\partial_\mu\chi_i F^i_{\nu\lambda}\nonumber\\
\end{eqnarray}
We can then integrate out $F^i_{\mu\nu}$ and arrive at a purely
scalar Lagrangian
\begin{equation}
S_{dual} = \int d^3x \sqrt{-g}\Big(
R - \frac{1}{2}e^{2\frac{n+1}{n}\alpha\rho}
 G^{ij}\partial_\mu\chi_i \partial^\mu\chi_j
-  {\rm Tr}(S_\mu S^\mu) -\frac{1}{2}\partial_\mu\rho\partial^\mu\rho\Big)
\label{scalarL}
\end{equation}
Let us consider the $n+1$-dimensional vielbein
\begin{equation}
\tilde e_{I}^{\ \overline J} =\left(\matrix{
e^{\alpha \rho} & e^{\alpha \rho}\chi_i\cr
0& e^{-\alpha\rho/n}e_i^{\ \overline j}\cr
}\right)
\label{Eenh}
\end{equation}
where $I,J=0,...,n$. The symmetric part
of $ e_{\overline J}^{\ \ K}\partial_\mu e_K^{\ \overline I}$ is
\begin{equation}
\tilde S_{\mu \ \overline J}^{\ \overline I} = \left(\matrix{
\alpha\partial_\mu \rho & \frac{1}{2}e^{\frac{n+1}{n}\alpha\rho}
e^{\overline i k}\partial_\mu\chi_k\cr
\frac{1}{2}e^{\frac{n+1}{n}\alpha\rho}e^{\ k}_{\overline j}\partial_\mu\chi_k & S_{\mu\ \overline j}^{\ \overline i}
-\frac{1}{n}\alpha\delta^{\overline i}_{\overline j}\partial_\mu\rho
}\right)
\end{equation}
Therefore we have
\begin{equation}
\tilde S_{{\mu \ \overline{J}}}^{\ \overline I}\tilde S^{{\mu\overline{J}}}_{\ \ \ \overline I}
=
S_{{\mu\ \overline j}}^{\ \overline i}S^{{\mu \overline j}}_{\ \ \ \overline i}
+\frac{1}{2}e^{2\frac{n+1}{n}\alpha\rho}G^{ij}\partial_\mu\chi_i\partial^\mu\chi_j
+ \frac{1}{2}\partial_\mu\rho\partial^\mu\rho
\end{equation}
Hence, after dualization,
the dimensionally reduced action can be written as
\begin{equation}
S_{dual}=\int d^3x \sqrt{-g}\Big(
R -{\rm Tr}(\tilde S_\mu \tilde S^\mu)\Big)
\label{dualgravity}
\end{equation}
This action is three-dimensional gravity coupled to an
$SL(n+1)/SO(n+1)$ coset, {\it. i.e.} a non-linear realization of
$SL(n+1)$ with local subgroup $SO(n+1)$.

Let us return to the dualized
Lagrangian (\ref{scalarL}).
We have already demonstrated that this Lagrangian can in fact be
written as an $SL(n+1)/SO(n+1)$ coset Lagrangian in terms
$\tilde e_I^{\ \ \overline J}$. However
we now wish to explain a simple procedure  for identifying the
enhancement of $SL(n)$ to $SL(n+1)$ that  will be important for  analyzing the
more complicated higher derivative terms that are the main subject of this paper.

The dualized Lagrangian (\ref{scalarL}) has a manifest $SL(n)$
symmetry. The  vielbein $e_i{}^{\overline j}$ of the internal torus
plays the role of the $SL(n)$ group element with local $SO(n)$
symmetry. As outlined in appendix B, this leads  to the Cartan forms
of $SL(n)$ out of which the ${\rm Tr}( S_\mu S^\mu)$ term in the
action is constructed. As such, we may write the vielbein as
\begin{equation}
e= e^{\sum_{\underline \alpha>0}\chi_{\underline\alpha}E_{\underline \alpha}}
\ e^{-\frac{1}{\sqrt{2}}\underline\phi \cdot \underline H}
\end{equation}
where $\underline H$ and $E_{\underline \alpha}$ are the Cartan sub-algebra and
positive root generators of $SL(n)$ respectively.
When evaluating the Cartan forms it does not matter what representation
of these generators we take as the calculation only depends on the structure constants of the Lie algebra.
However, to recover the vielbein
we need to consider the representation of $SL(n)$ that has highest weight
$\underline \lambda^{n-1}$ and also, as explained in  appendix C, its
dual which is a
representation with highest weight $\underline\lambda^1$.
If we choose as our basis of states of this latter
representation as $<j, \underline\lambda^1 | $, $j=1,...,n$, then the vielbein on the torus is given by
\begin{equation}
<j ,\underline\lambda^1| U(e)= \sum_ i <i,\underline\lambda^1 | e_i{}^{\overline j}
\end{equation}
where $U$ is the representation. We observe that (see equation
(\ref{app}) in appendix C) this object transforms only under local
rotations. Taking into account the transformation of the basis
vectors, one sees that the vielbein transforms  under local
rotations on its upper over-lined index and rigid $SL(n)$
transformations in the vector representation on its lower index. The
metric on the torus is then given by
\begin{equation}
G_{ij}= <i,\underline\lambda^1| U({\cal M})|j,\underline\lambda^1>
\end{equation}
where, as before, ${\cal M}=ee^\#$. Here the kets transform under
the Cartan twisted $\underline\lambda^{n-1}$ representation, which
is the
 $\underline\lambda^{1}$ representation.

Similarly we find that the inverse metric is given by taking the $SL(n)$
representation with highest weight
$\underline\lambda^{n-1}$
\begin{equation}
G^{ij}= <i, \underline\lambda^{n-1} |U({\cal M}^{-1})|j, \underline\lambda^{n-1}>
\end{equation}
As explained in appendix C,  we should take the bras to be the dual
of the Cartan twisted $\underline\lambda^{n-1}$ representation,
which is simply the $\underline\lambda^{n-1}$   representation.

Our goal here is to identify how the  $SL(n+1)$
roots arise in (\ref{scalarL}) from a more abstract point of view,
which  will generalize to the more complicated
 terms that occur in the higher derivative corrections.
The main   term in the action,  apart from the usual coset action for $SL(n)$, is
$G^{ij}\partial_\mu\chi_i \partial^\mu\chi_j $ which arises from the
dual graviphotons.
We will now see how this  term leads to the enhanced $SL(n+1)$ symmetry.
In particular we wish to show how the positive roots of
$SL(n+1)$, which are not roots of  $SL(n)$, arise from this term.

To proceed it is helpful to introduce different notations for
$n$ and $(n+1)$ dimensional vectors which arise in $SL(n)$ and $SL(n+1)$ respectively. We will denote the
former using a bar and the later using an arrow. In particular we will
write
\begin{equation}
\vec \phi = (\rho,\underline\phi)
\end{equation}
The roots $\underline\alpha$ of $SL(n)$
that appear in $S_\mu$ can be lifted to roots of $SL(n+1)$ simply
by taking
\begin{equation}
\vec \alpha = (0,\underline \alpha)
\end{equation}

To  evaluate $\partial_\mu\chi_i G^{ij}\partial_\mu\chi_j$
we note that $\partial_\mu \chi_i$ transforms as a vector under
the manifest $SL(n)$ symmetry that is in the representation with highest weight
$\underline \lambda^1$. As explained in appendix C,  we may write this term as
\begin{equation}
<\chi | U({\cal M}^{-1}) |\chi>
\end{equation}
In carrying out this step we have expressed $| \chi>$  in terms of
the basis states of the $\underline \lambda^1$ representation, that
is $|\chi>=\sum_i |i,\underline \lambda^1>$. Also as we are only
interested in the group theoretic structure of this term we have
suppressed the presence of space-time derivatives and other
unimportant factors. We have that
\begin{equation} {\cal M}^{-1} =
e^{-\sum_{\underline \alpha>0} \chi_{\underline\alpha}E_{-\underline
\alpha}} \ e^{\sqrt{2}\underline\phi \cdot \underline H}
\ e^{-\sum_{\underline \alpha> 0}\chi_{\underline\alpha}E_{\underline \alpha}}\nonumber\\
\end{equation}
and hence
\begin{eqnarray}
<\chi | U(M^{-1}) |\chi> &=& \sum _{w\ \in\ [\underline\lambda^1]}<w
| \ e^{\sqrt{2}\underline\phi \cdot \underline H} |w> +\ldots
\nonumber\\
&=& \sum _{w\ \in\ [\underline\lambda^1]}\ e^{\sqrt{2}\underline\phi
\cdot \underline w}<w|w>+\ldots\nonumber
\end{eqnarray}
where $[\underline\lambda^1]$ denotes the class of weights that
appear in the representation whose highest weight is
$\underline\lambda^1$ and the ellipsis denotes
$\chi_\alpha$-dependent terms which also have exponentials
containing the above weights. We have  used a basis for the
representation that consist of states which are eigenstates of
$\underline H$, that is are labeled by the weights of the
$\underline \lambda^1$ representation of $SL(n)$.

We are interested in the vectors that occur together with $\vec \phi$ in the exponential in the
$ \frac{1}{2}e^{2\frac{n+1}{n}\alpha\rho}
 G^{ij}\partial_\mu\chi_i \partial^\mu\chi_j $ term of the action (\ref{scalarL}). We find the vectors
\begin{equation}
\vec W  = (\sqrt{2}\frac{n+1}{n}\alpha, [\underline \lambda^{1}])
\label{Wvec}
\end{equation}
where $[\underline\lambda^{1}]$  denotes all the weights that occur
in the $SL(n)$ representation with highest weight
$\underline\lambda^1$. In particular  these are of the form of
$\underline \lambda^1$ minus a particular set of positive roots of
$SL(n)$.  As explained  in appendix C,   the  fundamental
representation of $SL(n)$ with highest weight $\underline
\lambda^{n-k}$ and lowest weight $\underline\mu^{n-k}$ is related to
the fundamental representation with highest weight $\underline
\lambda^{k}$ by $\underline\mu^{n-k}=-\underline \lambda^{k}$ .
Hence $[\underline \lambda^{1}]$ consists of the root string
beginning with $\underline\lambda^1$ and ending with
$-\underline\lambda^{n-1}$.

In the action of (\ref{scalarL}) we find  the roots of $SL(n)$ as
the coefficients of $\vec \phi$  in the exponentials
form the $Tr (S_\mu S^\mu) $ term. We also find the above set of vectors which arise from the
$ \frac{1}{2}e^{2\frac{n+1}{n}\alpha\rho}
 G^{ij}\partial_\mu\chi_i \partial^\mu\chi_j $ term.
The simplest way to realize that we just have the set of positive
roots of $SL(n+1)$ is to consider taking the lowest weight for
$[\underline\lambda^1]$ in (\ref{Wvec}) {\it i.e} replace
$[\underline\lambda^1]$ with $-\underline\lambda^{n-1}$. This is
just the vector
\begin{equation}
\vec \alpha_{n}=( \sqrt{2}\frac{n+1}{n}\alpha, -\underline
\lambda^{n-1} )
\end{equation}
whose scalar products are
\begin{eqnarray}
\vec \alpha_n\cdot \vec \alpha_n
&=&2\nonumber\\
\vec \alpha_n\cdot \vec\alpha_{n-1} &=& -1 \nonumber\\
\vec \alpha_n\cdot \vec\alpha_{a} &=& 0\qquad  a < n-1
\nonumber\\
\end{eqnarray}
Here we recognize the Cartan matrix of $SL(n+1)$.  We note that this
root ``attaches'' itself to the  $(n-1)$th node of the
 $SL(n)$ Dynkin diagram.
 Thus the simple roots of $SL(n+1)$ are given by
 \begin{equation}
 \vec\alpha_a=(0,\underline \alpha_a),\  a=1,\ldots , n-1;\  \  \vec\alpha_{n}=( \sqrt{2}\frac{n+1}{n}\alpha,
-\underline   \lambda^{n-1} )
\end{equation}
It is straightforward to see that the other vectors in (\ref{Wvec})
are positive integer combinations of the above simple roots since,
when selecting $\vec\alpha_{n}$, we took the lowest weight vector in
the $SL(n)$ representation. Therefore by construction all other
vectors are obtained from this one by the addition of  the positive
roots $\vec \alpha_a$ of $SL(n)$.

We observe that the fundamental weights $\vec\lambda^a$ of $SL(n+1)$
can be expressed in terms of the fundamental weights
$\underline\lambda^a$ of $SL(n)$ by \cite{Gaberdiel:2002db}
\begin{equation}
\vec\lambda^a = \left(\frac{\sqrt{2}\alpha}{n}a,\underline \lambda^a\right)\ a=1,...,n-1\qquad \vec\lambda^n =
(\sqrt{2}\alpha,\underline 0)
\end{equation}
Taking the above expression for the roots of $SL(n+1)$ it is easy to verify that these obey
the equation $\vec\alpha_a\cdot \vec\lambda^b=\delta_{a}^b$.

In summary we have observed that,
by looking for roots in the exponential terms,
we can  see an enhanced
$SL(n+1)/SO(n+1)$  coset structure in the dimensional reduction of the gravity when suitably dualized. This is an old result, but our purpose in this section has been to introduce a new technique for deriving this result that will prove very effective when analyzing the very complicated higher derivative terms.

\subsection{Higher Derivative Corrections}

Let us now turn our attention to an effective action for gravity
that includes higher derivative terms. One might have hoped that,
after compactification to three dimension and dualization, the
higher order terms can be written entirely in terms of the enhanced
$SL(n+1)/SO(n+1)$ coset Cartan form $\tilde S_\mu$. However this
cannot be the case. To see this we note that the form of the
curvature components shows that a generic higher derivative term
will be proportional to $e^{(2-l)\alpha\rho}$ where $l$ counts the
number of derivatives. At the lowest order $l=2$ and this factor
disappears, essentially because we are in Einstein frame. At a
general order $l$ the action has the form
\begin{equation}
S_l = \int d^Dx\sqrt{-\hat g}(\hat R)^\frac{l}{2}
\label{higheract}
\end{equation}
where we have suppressed the possible combinations of Lorentz
indices that can appear. Upon compactification this will lead to
terms of the form $e^{(2-l)\alpha\rho} (S_\mu S^\mu)^l$ with $l>2$
and these cannot be written in terms of $\tilde S_\mu$ alone since
the roots that appear in the latter, which  contain $\rho$
contributions, also come with $\partial\chi$-type terms and hence
are higher order in derivatives.

Therefore, since $\tilde S_\mu$ is determined entirely by the roots
of $SL(n+1)$, when we evaluate the higher derivative terms
we do not expect to  find only  the roots of $SL(n+1)$ appearing in the
exponential terms, as we did for the lowest order terms. Arguably the
next best thing is to hope that the exponential terms  will involve
the weights of $SL(n+1)$. Indeed we will now show that this is the case.

For most of this paper we will assume that the higher derivative
corrections only involve first order derivatives. It was shown in
\cite{Meissner} that the next-to-leading order corrections in the
effective action for gravity and the Bosonic String can always be
made to be first order in derivatives by a field redefinition.
Indeed one might expect that this is possible in general as a
well-defined perturbation expansion of some underlying microscopic
theory should not alter the degree of the equation of motion, which
would then require that additional initial data must be specified on
an initial value hypersurface. In other words we expect that in the
underlying quantum theory the Cauchy problem  does not have to be
reformulated at every order in perturbation theory (at least for a
certain choice of field variables). Furthermore the inclusion of
higher order derivative terms would complicate the dualization of
vector fields into scalars. In the presence of such terms one could
proceed by simply using the first order dualization prescription
however this would neglect many higher order contributions.
Nevertheless in section five we show that, assuming the same
dualization prescription that we use for second order equations of
motion,  the inclusion of terms involving more than one spacetime
derivative does not affect our conclusions.

From the expressions in the appendix ones see that the higher derivative
action, once compactified to three dimensions, will take the form
\begin{equation}
S = \int d^3 x e^{2\alpha\rho}L(e^{(\beta-2\alpha)\rho}F^i,
e^{-\alpha\rho}\partial\rho,e^{-\alpha\rho}S)
\end{equation}
where we have suppressed all Lorentz indices and dependence on the
three-dimensional metric for simplicity.
Next we dualize this Lagrangian by adding the Lagrange multiplier term
\begin{equation}
S = \int d^3 x e^{2\alpha\rho}L(e^{(\beta-2\alpha)\rho}F^i,
e^{-\alpha\rho}\partial\rho,e^{-\alpha\rho}S)
+  \int d^3x\epsilon^{\mu\nu\lambda}\chi_i\partial_\mu F^i_{\nu\lambda}
\end{equation}
Integrating the Lagrange multiplier term by parts and solving for $F^i$ one
finds
\begin{equation}
F^i = e^{-(\beta-2\alpha)\rho} f^i(e^{-\beta\rho}\partial\chi_i,e^{-\alpha\rho}\partial\rho,e^{-\alpha\rho}S)
\end{equation}
Substituting this back into the action leads to
\begin{equation}
S_{dual} = \int d^3 x \sqrt{-g}e^{2\alpha\rho}L_{dual}(e^{-\beta\rho}\partial
\chi_i,
e^{-\alpha\rho}\partial\rho,e^{-\alpha\rho}S)
\end{equation}
From this we see that a generic higher derivative term (\ref{higheract})
of order $l$ will lead to three-dimensional terms of the form
\begin{equation}
e^{2(1-r-s+\frac{t}{n})\alpha\rho}(\partial \rho)^{2r}(S)^{2s}(\partial \chi_i)^{2t}
\end{equation}
where the order is
\begin{equation}
l=2r+2s+2t
\end{equation}
We have used that fact that only even powers of
$\partial\rho$, $S$ and $\partial\chi_i$ can appear. To see this we
observe that since $x^i \to -x^i$ is a symmetry of the action we must
have that $A^i_\mu \to -A^i_\mu$ is a symmetry and hence
the dual action is an
even function of $\partial \chi_i$. We also note that if $e$  is a
choice of internal vielbein, {\it i.e.} an $n\times n$ matrix with
unit determinant, then so is $e^{-1}$. Therefore if $S$ is a solution
to the equations of motion so is $-S$ and hence the action must be an
even function of $S$. It then follows that the action must be an
even function of $\partial\rho$ since it must be even in derivatives.

We can now read off the  vectors that appear in the exponentials of
$\vec \phi$.  Using appendix C, and the analysis of the previous section,  we find the vectors
\begin{equation}
\vec W_{r,s,t} =
\Big(\sqrt{2}\Big(1-\frac{l}{2}+\frac{n+1}{n}t\Big)\alpha,s[\underline \theta ]
+ t[\underline\lambda^1]\Big)
\label{Wpure}
\end{equation}
We note that the coset representative $S_{\overline i}^{\ \overline
j}$ only contains positive $SL(n)$ roots and so only the positive
roots in the equivalence class $[\theta]$ actually appear. As above
$[\underline\lambda ]$ means the collection of weights that occur in
the representation with highest weight $\underline\lambda$. As a
result we note that the subscripts $(r,s,t)$ do not uniquely specify
the vectors that arise. The highest weight for the adjoint
representation is called $\underline\theta$ and in particular, for
$SL(n)$,
$\underline\theta=\underline\lambda^1+\underline\lambda^{n-1}$.
These contributions  arise from the factors of $S_\mu$ while those
proportional to $t$ arise from the graviphoton.

At order $l=2$, so that one of $(r,s,t)$ equals two with the others
vanishing, we recover the roots $\vec\alpha_a$, $a=1,..,n$ of
$SL(n+1)$ that we considered above. More generally we may write the
vector (\ref{Wpure}) as
\begin{equation}
\vec W_{r,s,t} =\vec W^H_{r,s,t}+ {\rm negative\ roots\ of\ the\ }
SL(n)\ {\rm subalgebra}
\end{equation}
where
 \begin{equation}
\vec W^H_{r,s,t}=\Big(\sqrt{2}\Big(1-\frac{l}{2}+\frac{n+1}{n}t\Big)\alpha,s\underline \theta
+ t\underline\lambda^1\Big)
\end{equation}
Recalling that $\underline \theta
=\underline\lambda^1+\underline\lambda^{n-1}$ one can show
that
\begin{equation}
\vec W^H_{r,s,t}= (1-2s-r)\vec \lambda^n+(t+s)\vec \lambda^1+s\vec
\lambda^{n-1}
\label{hhh}
\end{equation}
It follows that the vectors  that have arisen in the higher derivative terms belong to the weight lattice of
$SL(n+1)$.

Any weight can be expanded in terms of the fundamental weights $\vec
\lambda^a$ as $\vec \lambda = \sum c_a\vec\lambda^a $. We say that
$\vec\lambda$ is dominant if and only if $c_a \ge 0$ for all $a$.
A central theorem of Lie algebras asserts that the finite dimensional irreducible
representations of a Lie group are in one-to-one correspondence with
the dominant weights. Furthermore
the highest weight of an irreducible representation must be
dominant. The difference between any two weights of an
irreducible representation always gives a root. One can define the
highest weight to be the unique weight $\vec \Lambda^H$ such that
$\vec\Lambda^H+\vec\alpha$ is not a weight in the representation for
any positive root $\vec \alpha$. Similarly the lowest weight  $\vec
\Lambda^L$ such that $\vec\Lambda^H-\vec\alpha$ is not a weight in
the representation for any positive root $\vec \alpha$.

We now wish to identify what representation of $SL(n+1)$ these
weights belong to. If there they can be identified with the weights
of an irreducible representation then the highest weight must be
expressible as positive integers times the fundamental weights of
$SL(n+1)$. Examining equation (\ref{hhh}) ones see that we can only
take $r=s=0$ or $ r=1, s=0$. Taking the former choice we claim that
the highest weight is
\begin{equation}
\vec \lambda^{H} = \frac{l}{2}\vec \lambda^1 + \vec\lambda^n
\end{equation}
To prove  that this is indeed the highest weight of a representation
we must show that this weight, minus any other weight, is a sum of
positive roots. To this end we evaluate
\begin{eqnarray}
\vec \lambda^{H} -\vec W_{r,s,t}&=& s(2\vec \lambda^n-\vec
\lambda^{n-1})+r (\vec \lambda^n+\vec \lambda^1)\nonumber\\
&&+ {\rm positive\ roots\ of\ the\ } SL(n)\ {\rm subalgebra}\nonumber\\
\end{eqnarray}
Lastly we can show by direct computation that $\vec \lambda^1+\vec
\lambda^n$ and $2\vec\lambda^n-\vec\lambda^{n-1}$ have non-negative
and integral innerproducts with the fundamental weights. This
implies that they are positive roots. Therefore since $\vec
\lambda^{H} -\vec W_{r,s,t}$ is a sum of positive roots we conclude
that $\vec\lambda^H$ is indeed the highest weight. We note that this
calculation also shows that the difference between any two weights
that we find is indeed a root.

It is also instructive to identify the lowest weight $\vec\lambda^L$ that appears in the
effective action, in the sense that $\vec W_{r,s,t}-\vec \lambda^L$ is a positive
root for all $\vec W_{r,s,t}$. By a similar
argument we find
\begin{equation}
\vec\lambda^L =
(1-\frac{l}{2})\vec\lambda^n
\end{equation}
(recall that
only the positive roots in the equivalence class $[\theta]$ appear).
We recall that, for $SL(n+1)$, the lowest weight in the representation whose
highest weight is  $\vec\lambda^a$  is $-\vec\lambda^{n+1-a}$. Therefore
the lowest weight in the representation whose highest weight is
$\vec\lambda^H$ is actually $-\vec
\lambda^1-\frac{l}{2}\vec\lambda^n$. However we see that
\begin{equation}
\vec\lambda^L =-\vec \lambda^1-\frac{l}{2}\vec\lambda^n + \vec\theta
\end{equation}
where $\vec\theta= \vec\lambda^{1 }+\vec\lambda^{n}$ is the highest
weight associated to the adjoint representation. This result is not
surprising as one expects that the local $SO(n+1)$ symmetry has been
used to gauge away terms that are associated with the lowest
weights. This is consistent with the lowest order Lagrangian which
only contains positive roots because all the terms associated to the
negative roots have been gauged away by $SO(n+1)$.

In summary we have seen that the higher derivative corrections,
which apparently cannot be written just in terms of an
$SL(n+1)/SO(n+1)$ coset, are still organized by $SL(n+1)$, in the
sense that the weights of $SL(n+1)$ appear in the exponential. Thus
our result shows evidence for an $SL(n+1)$ symmetry in a full theory
of gravity and not just the low energy effective theory.

\section{M-theory}

Our final example is to consider the effective action of M-theory
whose lowest order effective action is
\begin{equation}
S = \int d^{11} x \sqrt{-\hat g}\Big(\hat R - \frac{1}{48} \hat
G_{\hat\mu\hat\nu\lambda\hat\rho}G^{\hat\mu\hat\nu\lambda\hat\rho}\Big)+\ldots
\label{Mact}
\end{equation}
where $G_{\hat\mu\hat\nu\lambda\hat\rho}$ is a closed 4-form and the
ellipsis denotes a Chern-Simons term that will not play a role here.
Since we are interested in compactifying this action on an
eight-torus our discussion also applies to the type II superstring
theories. Note that here $D=11$ and we wish to reduce to
three-dimensions so that $n=8$ and hence $\alpha = 2/3$.

Upon dimensional reduction to three-dimensions the 4-form
$\hat G$ leads to two the dynamical fields
\begin{equation}
\hat G_{\mu\nu ij} = F_{\mu\nu ij}\ ,\qquad
\hat G_{\mu ijk} = \partial_\mu\chi_{ijk}
\end{equation}
Therefore, using the results for pure gravity, a
generic higher derivative term, one reduced to three dimensions,
will be of the form
\begin{equation}
S = \int d^3 x e^{2\alpha\rho}L(e^{(\beta-2\alpha)\rho}F^i,
e^{-\alpha\rho}\partial\rho,e^{-\alpha\rho}S, e^{-(2\alpha+2\beta)\rho}F_{ij},
e^{-(\alpha+3\beta)\rho}\partial\chi_{ijk})
\label{Mactform}
\end{equation}
Just as before we must introduce Lagrange multipliers for the two types
of 2-form fields strengths
\begin{eqnarray}
S &=& \int d^3 x e^{2\alpha\rho}L(e^{(\beta-2\alpha)\rho}F^i,
e^{-\alpha\rho}\partial\rho,e^{-\alpha\rho}S, e^{-(2\alpha+2\beta)\rho}F_{ij},
e^{-(\alpha+3\beta)\rho}\partial\chi_{ijk})\nonumber\\
&& + \int d^3x\epsilon^{\mu\nu\lambda}\chi_{i}\partial_\mu
F^i_{\nu\lambda } + \int
d^3x\epsilon^{\mu\nu\lambda}\chi^{ij}\partial_\mu F_{\nu\lambda ij}
\nonumber\\
\end{eqnarray}
where $F^i_{\mu\nu}$ are the graviphoton field strengths.
Integrating the Lagrange multiplier terms by parts and solving for
$F^i$ and $F_{ij}$ one finds
\begin{eqnarray}
F^i_{\mu\nu} &=& e^{-(\beta-2\alpha)\rho}
f^i_{\mu\nu}(e^{-\beta\rho}\partial\chi_i,
e^{-\alpha\rho}\partial\rho,e^{-\alpha\rho}S,e^{-(\alpha+3\beta)\rho}\partial\chi_{ijk},
e^{2\beta\rho}\partial\chi^{ij})\nonumber\\
F_{\mu\nu ij} &=& e^{(2\alpha+2\beta)\rho} f_{\mu\nu ij}
(e^{-\beta\rho}\partial\chi_i,
e^{-\alpha\rho}\partial\rho,e^{-\alpha\rho}S,e^{-(\alpha+3\beta)\rho}\partial\chi_{ijk},
e^{2\beta\rho}\partial\chi^{ij})
\nonumber\\
\end{eqnarray}
Therefore the dual Lagrangian
takes the form
\begin{equation}
S_{dual}
= \int d^3 x e^{2\alpha\rho}L(e^{-\beta\rho}\partial\chi_i,
e^{-\alpha\rho}\partial\rho,e^{-\alpha\rho}S,e^{-(\alpha+3\beta)\rho}\partial\chi_{ijk},
e^{2\beta\rho}\partial\chi^{ij})
\end{equation}
A generic term will therefore be
\begin{equation}
e^{2\left(1-r-s+\frac{t}{8} -\frac{5}{8}p_1- \frac{4}{8}p_2\right)\alpha\rho}(\partial\rho)^{2r}(S)^{2s}(\partial\chi_i)^{2t}
(\partial \chi_{ijk})^{2p_1}(\partial \chi^{ij})^{2p_2}
\end{equation}
where the order is $l = 2r+2s+2t+2p_1+2p_2$.

The vectors that we encounter will be $8$-dimensional so we take
\begin{equation}
\vec \phi = (\rho,\underline \phi)
\end{equation}
and in this way we find
\begin{equation}
\vec W_{r,s,t,p_1,p_2}
= \Big(w,s[\underline\theta]+t[\underline\lambda^1]
+p_1[\underline\lambda^3]+p_2[\underline\lambda^{6}]\Big)
\end{equation}
with
\begin{equation}
w=\frac{2\sqrt{2}}{3}\Big(1-\frac{l}{2}+\frac{9}{8}t +\frac{3}{8}p_1 +\frac{6}{8}p_2\Big)
\end{equation}
Here $[\underline\theta]$ is one of the  positive roots that appears
as a weight in  the adjoint representation and
$[\underline\lambda^a]$ is one of the  $SL(8)$ weights that appears
in the  representation whose highest weight is
$\underline\lambda^a$. Note that $\vec W_{r,s,t,p_1,p_2} $ is not
uniquely determined by specifying the positive integers
$(r,s,t,p_1,p_2)$ due to the different choice of representative of
$[\underline\lambda^a]$.

In $\vec W_{r,s,t,p_1,p_2}$ the weights   $[\underline\lambda^1]$
arise just as they did for the case of pure gravity. However we
should comment on why $[\underline\lambda^3]$ and
$[\underline\lambda^6]$ have appeared. These also arise in the same
manner that $[\underline\lambda^1]$ does, only rather than coming
from evaluating $\partial_\mu \chi_i G^{ij}\partial^\mu\chi_j$ they
arise from evaluating terms involving $\partial_\mu \chi_{ijk}
G^{il}G^{jm}G^{ln}\partial^\mu\chi_{lmn}$ and $\partial_\mu
\chi^{ij} G_{ik}G_{jl}\partial^\mu\chi^{kl}$ respectively. The
discussion that we had for pure gravity still applies except that we
must change the representation that the fields transform under. In
the former case, since $\chi_{ijk}$ transforms in the 3-fold
anti-symmetric representation we find the weights $[\underline
\lambda^3]$ of this representation. For $\partial_\mu \chi^{ij}
G_{ik}G_{jl}\partial^\mu\chi^{kl}$  one first notes that the
$\epsilon_{ijklmnpq}$ symbol can be used to lower the indices so
that $\chi_{klmnpq} =\epsilon_{ijklnmpq}\chi^{ij}$ transforms in the
6-fold anti-symmetric representation. Hence it appears with the
weights $[\underline\lambda^6]$.

At lowest order $l=2$ and the action is given by
(\ref{Mact}). Upon dimensional reduction and dualization the
vectors that appear are
\begin{eqnarray}
\vec W_{0,1,0,0,0}
&=& \left(0,\underline \alpha\right)\nonumber\\
\vec W_{0,0,1,0,0}
&=& \left(\frac{3\sqrt{2}}{4},[\underline \lambda^1]\right)\nonumber\\
\vec W_{0,0,0,1,0}
&=& \left(\frac{\sqrt{2}}{4},[\underline \lambda^3]\right)\nonumber\\
\vec W_{0,0,0,0,1}
&=& \left(\frac{2\sqrt{2}}{4},[\underline \lambda^{6}]\right)\nonumber\\
\end{eqnarray}
where we have used the fact that $n=8$. The first line just consists
of the roots of $SL(n)$ whereas the second line  gives the same
roots that  we saw for pure gravity and which enhance $SL(n)$ to
$SL(n+1)$ however this root is no longer simple in this case. In
particular we consider the third line and take the lowest weight in
the representation with weights  $[\underline\lambda^3]$
\begin{equation}
\vec\alpha_8  = \left(\frac{\sqrt{2}}{4},-\underline \lambda^{5}
\right)
\end{equation}
It is straightforward to see that $\vec\alpha_8\cdot\vec\alpha_8=2$
and $\vec\alpha_8\cdot \vec\alpha_{5}=-1$ with the other
innerproducts vanishing. The other choices of $W_{0,0,0,1,0}$ are
obtained from $\vec \alpha_8$ by adding a positive root from $SL(n)$
and hence these are positive but not simple roots. One also sees
that none of the choices for $W_{0,0,0,0,1}$ and $W_{0,1,0,0,0}$ are
simple. Thus we have recovered the root diagram for $E_8$. To
summarize these roots are
\begin{equation}
\vec \alpha_a = (0,\underline\alpha_a)\ ,a=1,...,7\ \qquad
\vec\alpha_8  = \left(\frac{\sqrt{2}}{4},-\underline \lambda^{5}\right)
\end{equation}

The fundamental weights can be calculated to be
\begin{eqnarray}
\vec \lambda^a &=& \left(\frac{3\sqrt{2}}{4}a,\underline\lambda^a\right)\
  ,a=1,2,3,4\nonumber\\
\vec \lambda^a &=& \left(\frac{5\sqrt{2}}{4}(8-a),\underline\lambda^a\right)\
  ,a=5,6,7\nonumber\\
\vec\lambda^8  &=& (2\sqrt{2},\underline 0)\nonumber\\
\end{eqnarray}
and hence one finds that
\begin{eqnarray}
\vec W_{r,s,t,p_1,p_2}&=&
\left(\frac{1}{3}-\frac{l}{6}-s-p_1-p_2\right)\vec\lambda^8 +
(s+t)\vec \lambda^1 +p_1\vec \lambda^3+p_2\vec \lambda^6 +
s\vec\lambda^7 \nonumber\\
&& +{\rm negative\ roots\ of\ the\ }SL(8)\  {\rm subalgebra}\nonumber\\
\end{eqnarray}
Therefore we see that the condition for $\vec W_{r,s,t,p_1,p_2}$ to be a weight is
\begin{equation}
\frac{1}{3}-\frac{l}{6} \in {\bf Z} \quad \iff \quad l = 6g+2
\end{equation}
for some $g = 0,1,2,...$. Thus the conjecture that the higher derivative
terms involve the weights of $E_8$ implies that they can only come at
orders $l=2,8,14,...$, {\it i.e.} $R$, $R^4$,$R^7$.... This result is well known from quantum considerations
of type IIA string theory however here we see that it is also a consequence
of the existence of an enhanced coset structure arising in three dimensions.

It is instructive to identify the highest and lowest weights that
appear at a fixed order $l$. By inspection one sees that a
reasonable guess for highest weight is
\begin{equation}
\vec\lambda^H = -\frac{1}{6}(l-2)\vec\lambda^8 +\frac{l}{2}\vec\lambda^1
\end{equation}
To prove this we proceed as we did for pure gravity and note that
\begin{eqnarray}
\vec\lambda^H - \vec W_{r,s,t,p_1,p_2} &=&
s(\vec\lambda^8-\vec\lambda^7)+r\vec\lambda^1 +
p_1(\vec\lambda^8+\vec\lambda^1-\vec\lambda^3)
+p_2(\vec\lambda^8+\vec\lambda^1-\vec\lambda^3)\nonumber\\
&&
+{\rm positive\ roots\ of\ the\ }SL(8)\  {\rm subalgebra}\nonumber\\
\end{eqnarray}
One can show by explicit calculations that
$\vec\lambda^8-\vec\lambda^7$, $\vec\lambda^1$,
$\vec\lambda^8+\vec\lambda^1-\vec\lambda^3$ and
$\vec\lambda^8+\vec\lambda^1-\vec\lambda^3$ are all positive roots
of $E_8$. A similar calculation shows that the lowest weight, in the sense that
$\vec W_{r,s,t,p_1,p_2}-\vec\lambda^L$ is a positive root for all
$\vec W_{r,s,t,p_1,p_2}$, is
\begin{equation}
\vec\lambda^L = -\frac{1}{6}(l-2)\vec\lambda^8
\end{equation}
We observe that these are not
dominant weights and therefore they cannot be the highest weights of
a finite dimensional irreducible representation of $E_8$. Note that $E_8$ is special
in that the weight lattice and root lattice are the same. Curiously
one finds that $\vec \Lambda^H$ and $\vec \Lambda^L$ are positive
and negative roots respectively. We also see that
\begin{equation}
\vec\lambda^L = \vec\lambda^H -\frac{l}{2}\theta
\end{equation}
where $\vec\theta = \vec\lambda^1$ is the highest weight of the
adjoint representation of $E_8$. As with pure gravity we expect that some
weights have been gauged away using the local symmetry. Precisely how this happens
is not clear to us and the pattern of weights that we find deserves further study.

We note that there is a separate argument for the condition
that $l=6g+2$ based on the fact that the $M$-theory effective action
is the lift to eleven dimensions of the type IIA effective action.
In particular consider a higher derivative term in the M-theory
effective action of the form
\begin{equation}
S_l \sim \int d^{11}x \sqrt{-\hat g}(\hat R)^\frac{l}{2}
\end{equation}
The type IIA metric, in string frame, is related to the 11-dimensional
metric through
\begin{equation}
d\hat s^2 = e^{-2\Phi/3}ds^2 + e^{4\Phi/3}(dx^{11}+A_\mu dx^\mu)^2
\end{equation}
where $\Phi$ is the dilaton. Therefore, in ten-dimensions,
\begin{eqnarray}
S_l &\sim& \int d^{10}x \sqrt{-g}e^{-8\Phi/3}
(e^{2\Phi/3}(R)+\ldots)^\frac{l}{2}\nonumber\\
&\sim&\int d^{10}x \sqrt{-g}e^{-(8-l)\Phi/3} ((R)^\frac{l}{2}+\ldots)
\nonumber\\
\end{eqnarray}
However from string perturbation theory  we recall
that $g_s = e^\Phi$ is the string coupling and  that the action is
an expansion in $e^{2(g-1)\Phi}$ with $g=0,1,2,...$. Thus we see that
$-(8-l)/3= 2(g-1)$ {\it i.e.} $l = 6g+2$.

Finally we should comment on other terms that are known to arise in the
M-theory effective action and which do not naively take the form that we have
considered. At lowest order there is a Chern-Simons term
that only involves the three-form gauge field
\begin{equation}
S_{CS} = \int  \hat G \wedge \hat G \wedge \hat C_3
\end{equation}
Upon reduction to three-dimensions, and integration by parts,
this leads to a term the form
\begin{equation}
S_{CS} = \int d^{3}x\
\epsilon^{\mu\nu\lambda}\epsilon^{i_1...i_8}
F_{\mu\nu i_1i_2}\chi_{i_3i_4i_5}\partial_{\lambda}\chi_{i_6i_7i_8}
\end{equation}
The effect of this term is to alter the dualization formulae by
adding a $\chi \partial \chi$ term. In particular it leads to higher
order $\chi\partial\chi$ terms which we have consistently neglected
throughout this paper.

At the next-to-leading order, {\it i.e.} $l=8$, there is an anomaly term
of the form \cite{DuffLiuMinasian}
\begin{equation}
S_{I_8} = \int  I_8(\hat R)\wedge \hat C_3
\end{equation}
where $I_8(R)$ is an eight-dimensional topological class.
 Upon reduction to three-dimensions this leads to
terms of the form
\begin{eqnarray}
S_{I_8} &=& \int d^3x e^{3\alpha\rho}X_8(e^{-\alpha\rho}\partial, e^{(\beta-2\alpha)\rho}F^i,
e^{-\alpha\rho}\partial\rho,e^{-\alpha\rho}S)
e^{-(\alpha+3\beta)\rho}\chi_{ijk}\nonumber\\
&&+e^{3\alpha\rho}X'_8(e^{-\alpha\rho}\partial,e^{(\beta-2\alpha)\rho}F^i,
e^{-\alpha\rho}\partial\rho,e^{-\alpha\rho}S)e^{-(2\alpha+2\beta)\rho}A_{\mu ij}
\nonumber\\
\end{eqnarray}
where we have allowed for the fact that the curvature terms involve
higher order derivatives acting on the fields. Some terms, namely
those which contain an undifferentiated $\chi_{ijk}$, play a role
similar to the Chern-Simons term and introduce higher order
$\chi\partial\chi$ corrections into the dualization formulae which
we neglect. We believe that the remaining terms are those that
contain an extra spacetime derivative and can be rearranged using
integrated by parts into the form
\begin{eqnarray}
S_{I_8} &=& \int d^3x e^{2\alpha\rho}\Omega_8(e^{(\beta-2\alpha)\rho}F^i,
e^{-\alpha\rho}\partial\rho,e^{-\alpha\rho}S)
e^{-(\alpha+3\beta)\rho}\partial \chi_{ijk}\nonumber\\
&&+e^{2\alpha\rho}\Omega'_8(e^{(\beta-2\alpha)\rho}F^i,
e^{-\alpha\rho}\partial\rho,e^{-\alpha\rho}S)e^{-(2\alpha+2\beta)\rho}F_{ ij}
\nonumber\\
\end{eqnarray}
Indeed this must be the case if this correction preserves the second
order nature of the equations of motion. However these terms are of
the type that we have already considered in (\ref{Mactform}),
although it is important to note that any internal
indices are not contracted using the metric
but rather an $\epsilon$-symbol and this will affect the calculation
of the weights. In any case we believe that the vectors which arise as
coefficients of the scalar fields in these terms are consistent with
the pattern of $E_8$ weights that we found above.

\section{The Bosonic String}

Our next example is the Bosonic
String in $D$-dimensions.
At lowest order in derivatives the effective action is
\begin{equation}
S = \int d^D x \sqrt{-\hat g}\Big(\hat R - \frac{1}{2}\partial_\mu \phi
\partial^\mu \phi -\frac{1}{12} e^{\sqrt{\frac{8}{D-2}} \phi}
\hat H_{\hat\mu\hat\nu\lambda}\hat H^{\hat\mu\hat\nu\lambda}\Big)
\label{BSact}
\end{equation}
where $\hat H$ is a closed 3-form and the scalar $\phi$
is related to the string theory dilaton $\Phi$ by
$\phi=-\sqrt{8/(D-2)}\Phi$. Since we will reduce to three-dimensions
we will take $n=D-3$ and hence $\sqrt{8/(D-2)} =\sqrt{8/(n+1)}$.
Upon dimensional reduction to three-dimensions the three form $\hat
H$ leads to the dynamical fields
\begin{equation}
\hat H_{\mu\nu i} = G_{\mu\nu i}\ ,\qquad
\hat H_{\mu ij} = \partial_\mu\chi_{ij}
\end{equation}

A generic higher derivative term, one reduced to three-dimensions,
will be of the form
\begin{equation}
S = \int d^3 x e^{2\alpha\rho}L(e^{(\beta-2\alpha)\rho}F^i,
e^{-\alpha\rho}\partial\rho,e^{-\alpha\rho}S,
e^{-(2\alpha+\beta)\rho}G_i,
e^{-(\alpha+2\beta)\rho}\partial\chi_{ij},\phi)
\end{equation}
Before proceeding to dualize the electromagnetic field strengths as
we did above  we note that the most general higher derivative action
will contain arbitrary exponentials of $\phi$ that multiply the
electromagnetic field strengths. This complicates the
dualization and we have been unable to find tractable formulae for
the general term in this case, as we did  for pure
gravity and M-theory. Therefore in this section we will
simply use the dualization formulae that arise at the lowest order.
In particular after reducing (\ref{BSact}) to three dimensions we add
the terms
\begin{equation}
\int d^3x\epsilon^{\mu\nu\lambda}\chi_i\partial_\mu F^i_{\nu\lambda}
+ \int d^3x\epsilon^{\mu\nu\lambda}\psi^i\partial_\mu G_{\nu\lambda
i}
\end{equation}
where $F^i_{\mu\nu}$ is the graviphoton field strength. Integrating
out the field strengths $F^i_{\mu\nu}$ and $G_{\mu\nu i}$ we find
\begin{eqnarray}
F^i_{\mu\nu} &=&\frac{1}{2}\epsilon_{\mu\nu\lambda}G^{ij}
e^{2(\alpha-\beta)\rho}\partial^\lambda\chi_j\nonumber\\
G_{\mu\nu i} &=&\frac{1}{2}
\epsilon_{\mu\nu\lambda}G_{ij}e^{2(\alpha+\beta)\rho}
e^{-\sqrt\frac{8}{n+1}\phi}\partial^\lambda\psi^i\nonumber\\
\label{lowestdual}
\end{eqnarray}
In general these equations will be corrected by higher order terms.
Therefore if we continue by substituting these lowest order
expressions into the full effective action we will in fact be
neglecting many additional terms. However the terms that we find in
this manner would still be present even if we could somehow perform
a complete treatment. Therefore we will proceed with the knowledge
that our results only capture a subset of all the terms that appear.
However we do not believe that our conclusions would be altered
by a more complete treatment.

To continue we observe that a generic term  in the effective action
will be of the form
\begin{equation}
e^{u\phi}e^{2\left(1- r-s-\frac{2n+1}{n}t -\frac{n-2}{n}q_1 -
\frac{2n-1}{n}q_2\right)\alpha\rho}(\partial\rho)^{2r}(S)^{2s}(F^i)^{2t}
(\partial \chi_{ij})^{2q_1}(G_i)^{2q_2}(\partial \phi)^{2v}
\label{Bgenterm}
\end{equation}
After dualization, using the lowest order expressions
(\ref{lowestdual}) this term takes the form
\begin{eqnarray}
e^{\left(u-2\sqrt{\frac{8}{n+1}}q_2\right)\phi}e^{2w\alpha\rho}
(\partial\rho)^{2r}(S)^{2s}(\partial\chi_i)^{2t} (\partial
\chi_{ij})^{2q_1}(\partial\psi^i)^{2q_2}(\partial \phi)^{2v}
\end{eqnarray}
where
\begin{equation}
w =1- r-s+\frac{1}{n}t -\frac{n-2}{n}q_1 -\frac{1}{n}q_2
\end{equation}
and the order is $l = 2r+2s+2t+2q_1+2q_2 + 2v$.

Note that since we also have the a dependence on the dilaton the
vectors that we find will have two additional entries as
compared to the weights of $SL(n)$. Therefore in this section we
define
\begin{equation}
\vec\phi = (\phi,\rho,\underline \phi)
\end{equation}
and hence with this notation we obtain
\begin{eqnarray}
\vec W_{r,s,t,q_1,q_2} &=& \Big(\frac{u}{\sqrt{2}} -
\frac{4q_2}{\sqrt{n+1}},\sqrt{2} \left(1-\frac{l}{2}+\frac{n+1}{n}t
+\frac{2}{n}q_1 +\frac{n-1}{n}q_2\right)\alpha,\nonumber\\
&& \ \ \ \ \ \ \ \ \ \ \ \ \ \ \ \ \ \ \ \ \ \ \ \ \ \
s[\underline\theta] +
t[\underline\lambda^1]+q_1[\underline\lambda^2]+q_2[\underline\lambda^{n-1}]\Big
)
\nonumber\\
\end{eqnarray}
Once again $[\underline\lambda]$ represents any of the weights that
appear in a representation whose highest weight is
$\underline\lambda$ and $\underline
\theta=\underline\lambda^1+\underline\lambda^{n-1}$ is the highest
weight  of the adjoint representation. We should comment on why
$[\underline\lambda^2]$ and $[\underline\lambda^{n-1}]$ appear.
However these arise in the same way that $[\underline\lambda^3]$ and
$[\underline\lambda^6]$ did in the previous section on M-theory.

At order $l=2$ find
\begin{eqnarray}
\vec W_{0,1,0,0,0} &=& \Big(0,0,[\underline\theta]\Big )\nonumber\\
\vec W_{0,0,1,0,0} &=& \Big(0,\sqrt{2}\frac{n+1}{n}\alpha,[\underline\lambda^1]\Big )\nonumber\\
\vec W_{0,0,0,1,0} &=& \Big(\sqrt{\frac{4}{n+1}},2\sqrt{2}\frac{1}{n}\alpha,[\underline\lambda^2]\Big )\nonumber\\
\vec W_{0,0,0,0,1} &=&
\Big(-\sqrt{\frac{4}{n+1}},\sqrt{2}\frac{n-1}{n}\alpha,[\underline\lambda^{n-1}]\Big
)\nonumber\\
\end{eqnarray}
where we have used the fact that at lowest order $u = 0 ,
\sqrt{\frac{8}{n+1}}$. In particular
\begin{eqnarray}
\vec \alpha_a &=& \Big(0,0,\underline\alpha_a\Big )\qquad a=1,...,n-1\nonumber\\
\vec \alpha_n &=& \Big(\sqrt{\frac{4}{n+1}},2\sqrt{2}\frac{1}{n}\alpha,-\underline\lambda^{n-2}\Big )\nonumber\\
\vec \alpha_{n+1} &=&
\Big(-\sqrt{\frac{4}{n+1}},\sqrt{2}\frac{n-1}{n}\alpha,-\underline\lambda^1\Big
)\nonumber\\
\end{eqnarray}
One can show that $\vec W_{0,0,1,0,0}$ is not simple,
{\it i.e.} it can be expressed as a linear combination of the
$\vec\alpha_a$  $a=1,...,n+1$ with integer
coefficients. The roots $\vec\alpha_a$ $a=1,...,n-1$ are those of
$SL(n)$. In addition
we find $\vec\alpha_n$ and $\vec\alpha_{n+1}$. Each has length
$\sqrt{2}$ and
the only other non-vanishing innerproducts are
$\vec\alpha_n\cdot \vec \alpha_{n-2}=\vec\alpha_{n+1}\cdot\vec \alpha_{1}=-1$.
Thus we have recovered the simple roots of $O(n+1,n+1)$.

To proceed we note that the fundamental weights are
\begin{eqnarray}
\vec \lambda^a &=& \left(\frac{a-1}{\sqrt{n+1}},\sqrt 2\alpha {(n
+a)\over n}, \underline\lambda^a\right)\
   ,a=1,\ldots , n-2\nonumber\\
\vec \lambda^{n-1} &=& \left(
\frac{1}{2}\frac{n-3}{\sqrt{n+1}},\sqrt
2\alpha {(n-1)\over n}, \underline\lambda^{n-1}\right)\nonumber\\
\vec \lambda^n &=& \left(\frac{1}{2}\frac{n-1}{\sqrt{n+1}},\sqrt
2\alpha , \underline 0
\right)\nonumber\\
\vec \lambda^{n+1} &=& \left(-\frac{1}{\sqrt{n+1}} ,\sqrt
2\alpha , \underline 0 \right)\nonumber\\
\end{eqnarray}
It follows that
\begin{eqnarray}
\vec W_{r,s,t,q_1,q_2} &=& (s+t)\vec \lambda^1 + q_1\vec \lambda^2
+(s+q_2)\vec\lambda ^{n-1}+w_n\vec\lambda^n +
w_{n+1}\vec\lambda^{n+1}
\nonumber\\
&& +{\rm negative\ roots\ of\ the\ }SL(n)\  {\rm subalgebra}\nonumber\\
\end{eqnarray}
where
\begin{eqnarray}
w_n &=&\frac{\sqrt{2}\sqrt{n+1}u + 2-l-s(n+1) -4q_1- q_2(n+5)}{n+1}\nonumber\\
w_{n+1} &=&1-\frac{l}{2}-2s-q_1 -w_{n}\nonumber\\
\end{eqnarray}
Thus we see that $\vec W_{r,s,t,q_1,q_2}$ will be a weight if and
only if $w_n,w_{n+1}\in {\bf Z}$. We note that $w_{n+1}$ is an
integer if and only if $w_n$ is an integer so that we arrive at the
condition
\begin{equation}
\frac{\sqrt{2}\sqrt{n+1}u + 2-l-s(n+1) -12q_1- q_2(n-3)}{n+1} \in
{\bf Z}
\label{Bintcond}
\end{equation}

To illuminate the physical content of this condition we observe that
the string metric $\hat g^S_{\hat\mu\hat\nu}$ is related to the
Einstein metric through $\hat
g^E_{\hat\mu\hat\nu}=e^{-\frac{4}{n+1}\Phi}\hat
g^S_{\hat\mu\hat\nu}$ with $\phi =-\sqrt{\frac{8}{n+1}}\Phi$. Thus
in string frame the general term (\ref{Bgenterm}) comes from a term
of the form
\begin{equation}
e^{2(g-1)\Phi}\sqrt{-\hat g^S} (\hat R)^{r+s+t}(\hat
H)^{2q_1+2q_2}(\partial \Phi)^{2v}
\end{equation}
with
\begin{equation}
g-1=\frac{2(r+s+t)+6q_1+6q_2+2v-(n+3)-\sqrt{2}\sqrt{n+1}u}{n+1}
\end{equation}
Such a term arises in string perturbation theory at genus $g$ and
hence we must identify $g$ with a non-negative integer. Solving for
$u$ as a function of $r,s,t,q_1,q_2,v$ and $g$ we can substitute
back into (\ref{Bintcond}) and find
\begin{eqnarray}
\vec W_{r,s,t,q_1,q_2} &=& (s+t)\vec \lambda^1 + q_1\vec \lambda^2
+(s+q_2)\vec\lambda ^{n-1}
\nonumber\\
&&-(g+s+q_2)\vec\lambda^n+(1+g+q_2-q_1-s-\frac{l}{2})\vec\lambda^{n+1}
\nonumber\\
&& +{\rm negative\ roots\ of\ the\ }SL(n)\  {\rm subalgebra}\nonumber\\
\end{eqnarray}
Thus the condition that we find weights of the enhanced coset is
equivalent to the statement that the higher order terms arise from
string perturbation theory on a genus $g$ surface. We note that all
these weights have a negative coefficient of $\vec\lambda^n$ when $l>2$ and
hence cannot be dominant. Therefore they cannot be identified with
the highest weights of a finite-dimensional irreducible representation of $O(n+1,n+1)$.
Although we recall that we have only used the lowest order
dualization procedure and hence that we have neglected many terms
and their associated weights.

We would like to comment futher on the nature of the terms that we
have ignored by taking the lowest order dualization formulae. Since
the higher order corrections to the dualization formulae can be
viewed as a power series in $e^{2\Phi}$ their inclusion will only
lead to corrections in the dualized Lagrangian which are also a
power series in $e^{2\Phi}$. Thus we expect that they will  also be
consistent with the condition that the vectors that we obtain are
weights of the $O(n+1,n+1)$ symmetry if and only if they arise from
a given order of string perturbation theory.

\section{Weights of Terms With Two or More Spacetime
Derivatives Acting on a Field.}

Although we have indicated that one might not expect to find higher
derivative terms that have more than one spacetime derivative acting
on a field, we will now examine what weights correspond to such
contributions. Let us first consider the case of pure gravity.
Examining (\ref{Rcomp}) of appendix A we find that in the Riemann
tensor there  are terms that have more than one spacetime derivative
acting on a field, {\it i.e.} $\partial F^{i}$, $(\partial S) $ and
$(\partial^2\rho)$. Rather than compute the weights of such terms
directly, it is simpler to compute the change of weight when we swap
the term $ F_{\mu\nu i} F^{\mu\nu i}$ by $\partial^2$. This
preserves the number of spacetime derivatives. Of course  the latter
factor acts on some other fields but for the purpose of deriving the
change in the weight one does not have to know which factor it acts
on. The term $F_{\mu\nu i}F^{\mu\nu i}$ leads to the  weight
$({\sqrt 2 \alpha \over n}, -[\underline \lambda^{n-1}])$ while
$\partial^2$ leads to $(-\sqrt 2 \alpha,0)$. Hence swapping the
former term by the latter leads to the change in weight
\begin{eqnarray}
\vec\Delta W &=&(-\sqrt 2\alpha{n+1\over n},
[\underline\lambda^{n-1}])\nonumber\\
&=&(-\sqrt 2\alpha{n+1\over n},
\underline\lambda^{n-1}) +{\rm negative\ root\ of \ the }\ SL(n)\ {\rm
subalgebra}\nonumber\\
&=& - \vec\alpha_n+{\rm negative\ root\ of \ the }\ SL(n)\ {\rm
subalgebra}
\end{eqnarray}
Thus $\vec\Delta W$ is a negative $SL(n+1)$ root. Since the higher derivative term that
includes $ F_{\mu\nu i} F^{\mu\nu i}$ leads to an $SL(n+1)$ weight,
we must conclude that including the above term also leads to
$SL(n+1)$ weights. Thus all terms that arise in a correction
constructed from the Riemann tensor lead to weights of $SL(n+1)$.

We now consider the case of M-theory. So far we have only considered
higher derivative terms that are polynomial in the Riemann tensor
and the four form field strength. The second derivatives that were
discussed above for gravity also arise in M-theory. However the
$SL(9)$ algebra of pure gravity compactified from eleven dimensions
to three dimensions is a subalgebra of $E_8$.
In particular one can show that, in terms of $E_8$ roots
\begin{equation}
\vec \Delta W = -3\vec\alpha_8 +{\rm negative\ root\ of \ the }\ SL(
8)\ {\rm
subalgebra}
\end{equation}
Thus this change will take weights of $E_8$ to other weights of $E_8$.
However, the higher derivative
terms that arise from the quantum corrections also involve
derivatives of the four form field strength. We will now discuss
these terms. Rather than work out all such terms and their
corresponding weights it is simpler to work out what is the change
of the weight when we replace the term $F_{\mu ijk}F^{\mu ijk}$ by
$\partial^2$. As remarked above,  the latter factor acts on some
other field strength, but for the purpose of deriving the change in
the weight one does not have to know which factor it acts on. The
$F_{\mu ijk}F^ {\mu ijk}$ factor gives rise to the
$e^{-2(\alpha+3\beta) \rho\cdot \phi}e^{\sqrt 2 \underline\lambda^3
\cdot \underline\phi} =e^{-{5\over 4}\alpha\rho}e^{\sqrt 2
\underline\lambda^3 \cdot \underline\phi}$ while $\partial^2$ just
comes with $e^{-2\alpha\rho}$. Hence the change in weight due to
replacing the former factor by the latter factor is
\begin{eqnarray}
\vec\Delta W^{(1)} &=&(-{\sqrt 2\over 4},
-[\underline\lambda^3])\nonumber\\
&=&(-{\sqrt 2\over 4},\underline\lambda^5)
+{\rm negative\ root\ of \ the }\ SL(8)\ {\rm
subalgebra}
\nonumber\\
&=&-\vec\alpha_8
+{\rm negative\ root\ of \ the }\ SL(8)\ {\rm
subalgebra}
\nonumber\\
\end{eqnarray}
Since this is a negative $E_8$ root we see that the replacement will
indeed change a weight into a weight.

One can also find the change in the weight that is induced by
swapping the term $F_{\mu\nu ij} F^{\mu\nu ij}$ with $\partial^2$.
Using the same techniques one finds that the change is given by
\begin{eqnarray}
\vec\Delta W^{(2)}&=&(-{2\sqrt 2\over 4}, -[\underline\lambda^6])\nonumber\\
&=&(-{2\sqrt 2\over 4},\underline\lambda^2) +{\rm negative\ root\ of
\ the }\ SL(8)\ {\rm subalgebra}
\nonumber\\
&=&\vec\lambda^2-\vec\lambda^8+{\rm negative\ root\ of \ the }\
SL(8)\ {\rm subalgebra}
\nonumber\\
&=&-\vec\alpha_8+{\rm negative\ root\ of \ the }\ SL(8)\ {\rm
subalgebra}\nonumber\\
\end{eqnarray}
and again we find that the replacement takes one weight to another.

Hence if we include these terms where two or more spacetime
derivatives act on the fields then we find that they also lead to
weights of $E_8$. In particular the replacement $F^2\to \partial^2$
always changes the weight by a negative root of the enhanced
symmetry algebra.

It is important to note that in carrying out these steps we have
assumed that the dualisation is the same as that found in the rest
of the paper. In particular this is obtained by considering only
terms that have only a single derivative acting on each field.  It
is not clear to us how the dualization procedure can be extended to
include terms with additional derivatives. It is likely that the
dualization procedure will be changed by the presence of these new
terms, however, it will still contain the original dualization
prescription as the local piece.

\section{Conclusions}

In this paper we have considered the higher derivative terms that
arise for pure gravity,  M-theory  and the Bosonic string when
dimensionally reduced to three dimensions. We have derived the
general dependence of the terms in the effective action on the
diagonal metric components (and dilaton when present)
and shown that these are given by the
weights of the enhanced symmetry group.
More precisely the diagonal metric components are associated
with the Cartan subalgebra of the symmetry group $G$ whereas the
off diagonal components give rise to scalars $\chi_{\underline\alpha}$
that are associated to raising operators $E_{\underline\alpha}$
in the Lie algebra. For the lowest order term in
the effective action this procedure leads to the positive roots and
hence uniquely identifies  the enhanced symmetry algebra.

In more detail we found that in the case of gravity, dimensionally
reduced on a $n$-torus to three dimensions, all the higher
derivative terms lead to the appearance of weights of $SL(n+1)$. For
the M-theory we find that only higher derivative terms of the form
$R^{3k+1}+\ldots $ lead to weights of $E_8$ while  other higher
derivative terms the vectors that arise have no interpretation in
terms of the $E_8$ Lie algebra. For the Bosonic string we found that
the weights of $O(n+1,n+1)$ occur only when the terms in the
effective action can be identified as arising at a particular genus
in the string perturbation expansion. Thus the higher derivative
terms for which weights appear are just those that are expected
based on arguments using the underlying string theory. Alternatively
one could view the appearance of weights as a way of predicting
which higher derivative terms should arise.

The appearance of weights for  precisely for those higher derivative terms
that are expected to arise from the underlying quantum theory
provides strong evidence for the existence of some form of the enhanced symmetry
in the fundamental theory.
If we had found that the vectors contained within the higher derivative terms
had no interpretation in terms of the symmetry algebra of the low energy effective action then one
would be led to conclude that the U-duality conjectures \cite{HT} were incorrect.
The same would apply to the $E_{11}$ conjecture \cite{West:2001as} for the eleven dimensional
theory as its dimensional reduction must possess the residual  symmetry.  Thus
the appearance of weights can be seen as support for both of these conjectures.
However, it might also tell us something about how one should think about
of these conjectures beyond the low energy theory. Fortunately, in the analysis
carried out in this paper one is discussing only a finite dimensional subalgebra
of the full $E_{11}$ symmetry and so the questions raised by this paper may be
easier to resolve.

We also note that the existence of the enhanced symmetry in three
dimensions is responsible for the fact that, upon further
dimensional reduction to two-dimensions, the symmetry of the lowest
order effective action becomes
infinite-dimensional (for example see
\cite{Geroch:1972yt,Nicolai,Julia:1996nu,Mizoguchi}). Thus our
results also support the conjectures that, at least in
two-dimensions, the full quantum theory possess an
infinite-dimensional symmetry.

This paper has been somewhat phenomenological in nature. In
particular we have observed a pattern in the higher derivative terms
which is associated to the enhanced symmetry group. However we have
not explained why this occurs. Therefore it is of interest to
understand the meaning of our results and in particular demonstrate
that the enhanced group really is a symmetry. The most naive
expectation is that the higher derivative terms can be expressed in
terms of the group element of the non-linear realization that appears at lowest order.
However at first sight this would not seem to be the case.
It could be that
the enhanced symmetry does not act simply once higher derivative terms are
included and in particular requires an order by order modification of the
group action.
It is also possible that the symmetry algebra becomes
enlarged in the presence of the higher derivative terms to
incorporate the new representations that arise. We hope to report on
these points in the near future.

\section*{Acknowledgements}

We would like to thank Marc Henneaux for discussion on the BKL limit
and Andrew Pressley for discussions on group theory.
NL would like to thank the KITP, Santa Barbara and
PW would like to thank the CECF, Valdivia, Chile for their hospitality
where part of this work was carried out.
PW is supported by a PPARC
senior fellowship PPA/Y/S/2002/001/44.
This work was in addition supported in part by the PPARC grant
PPA/G/O/2000/00451 and the EU Marie Curie research training work
grant HPRN-CT-2000-00122.

\section*{Appendix A}

Our compactification ansatz is
\begin{equation}
d\hat s^2 = e^{2\alpha \rho}ds^2 + e^{2\beta\rho}G_{ij}
(dx^i + A^i_\mu dx^\mu)
(dx^j + A^j_\mu dx^\mu)
\end{equation}
where a hat denotes a $D$-dimensional quantity.
In this appendix we will quote results for a general compactification
from $D$-dimensions to $d$-dimensions.
We will use the indices
$\mu,\nu=0,1,2,..,d-1$ and $i,j = d+1,...,D-1$ and an overlined
index refers to the tangent frame.  The internal metric $G_{ij}$ is
constrained to have unit determinant (so that $\rho$ alone determines
the volume of the internal torus). We will also use $n=D-d$ to denote
the dimension of the internal torus.

A vielbein frame for this compactification is
\begin{eqnarray}
\hat e^{\overline \nu} &=& e^{\alpha\rho}e_\mu^{\ \ \overline \nu}dx^\mu\nonumber\\
\hat e^{\overline i} &=& e^{\beta\rho}e_j^{\ \ \overline i}
(dx^j+ A^{j}_\mu dx^\mu)
\end{eqnarray}
 where $e_\mu^{\ \ \overline \nu}$ and
$e_j^{\ \ \overline i} $ are vielbein frames for
$g_{\mu\nu}$ and $G_{ij}$ respectively.

The spin connection is
\begin{eqnarray}
\hat \omega^{\overline i}_{\ \ \overline j} &=&-e^{-\alpha\rho}Q_{\overline
\mu\  \overline j}^{\  \overline i}\hat e^{\overline\mu}\nonumber\\
\hat \omega^{\overline i}_{\ \ \overline \nu} &=&\beta e^{-\alpha\rho}\partial_{\overline\nu}\rho
\hat e^{\overline i} +\frac{1}{2}e^{(\beta-2\alpha)\rho}F^{\overline i}_{\overline\nu\overline\lambda }
\hat e^{\overline\lambda}+e^{-\alpha\rho}S_{\overline\nu\ \overline j}^{\ \overline i}\hat e^{\overline j}\nonumber\\
\hat \omega^{\overline \mu}_{\ \ \overline \nu} &=& \omega^{\overline \mu}_{\ \ \overline \nu}
+\alpha\partial_{\overline \nu}\rho e^{-\alpha\rho}\hat e^{\overline \mu}
-\alpha\partial^{\overline \mu}\rho e^{-\alpha\rho}\hat e_{\overline \nu}
-\frac{1}{2}e^{(\beta-2\alpha)\rho}F^{\ \overline \mu}_{\overline i\ \ \overline\nu}\hat e^{\overline  i}\nonumber\\
\end{eqnarray}
Here we have introduced
\begin{equation}
F^i_{\mu\nu} = \partial_\mu A_\nu^i - \partial_\nu A_\mu^i
\end{equation}
and
\begin{eqnarray}
e_{\overline j}^{\ \  k} \partial_\mu e_k^{\ \ \overline i}
&=& S_{\mu\  \overline j}^{\  \overline i} + Q_{\mu\  \overline j}^{\  \overline i}
\end{eqnarray}
where
$Q_\mu^{(\overline {ij})} = S_\mu^{[\overline {ij}]}=0$.
This split leads to the identities
\begin{eqnarray}
\partial_\mu Q_\nu-\partial_\nu Q_\mu - [Q_\mu,Q_\nu]-[S_\mu,S_\nu]&=&0\nonumber\\
\partial_\mu S_\nu-\partial_\nu S_\mu - [Q_\mu,S_\nu]+[Q_\nu,S_\mu]&=&0\nonumber\\
\label{constraints}
\end{eqnarray}

We can now calculate the Riemann curvature terms to be
\begin{eqnarray}
\hat R_{\overline{klij}}&=&e^{-2\alpha\rho}\Big(
S_{\overline{\nu il}}S^{\overline\nu}_{jk}-S_{\overline{\nu jl}}S^{\overline\nu}_{ik}
+\beta\partial^{\overline\nu}\rho(S_{\overline{\nu il}}\delta_{\overline{kj}}
-S_{\overline{\nu ik}}\delta_{\overline{lj}}
+S_{\overline{\nu jk}}\delta_{\overline{li}}
-S_{\overline{\nu jl}}\delta_{\overline{ik}})\nonumber\\
&&+\beta^2(\partial\rho)^2(\delta_{\overline{kj}}\delta_{\overline{li}}-\delta_{\overline{ki}}\delta_{\overline{lj}})\Big)\nonumber\\
\hat R_{\overline{\mu kij}}&=&\frac{1}{2}e^{(\beta-3\alpha)\rho}\Big(
S^{\overline\nu}_{\overline{ik}}F_{\overline{j\nu\mu}}
-S^{\overline\nu}_{\overline{jk}}F_{\overline{i\nu\mu}}
-\beta\partial^{\overline\nu}\rho F_{\overline{i\nu\mu}}\delta_{\overline{jk}}
+\beta\partial^{\overline\nu}\rho F_{\overline{j\nu\mu}}\delta_{\overline{ik}}
\Big)\nonumber\\
\hat R_{\overline{\mu\nu ij}}&=&2e^{-2\alpha\rho}\Big(-\nabla_{[\overline \mu}Q_{\overline\nu]\overline{ij}}+(Q_{[\overline\mu}Q_{\overline\nu]})_{\overline{ij}}+\frac{1}{4}e^{2(\beta-\alpha)\rho}
(F_{\overline i}F_{\overline j})_{[\overline{\mu\nu}]}\Big)\nonumber\\
\hat R_{\overline{\mu i \nu j}}
&=& e^{-2\alpha\rho}\Big((2\alpha\beta-\beta^2)\partial_{\overline \mu}\rho\partial_{\overline \nu}\rho\delta_{\overline{ij}}
-\beta \nabla_{\overline \mu}\partial_{\overline \nu}\rho \delta_{\overline{ij}}
-\alpha\beta (\partial\rho)^2\eta_{\overline{\mu\nu}}\delta_{\overline{ij}}
\nonumber\\
&&
-\frac{1}{4} e^{2(\beta-\alpha)\rho}(F_{\overline i}F_{\overline j})_{\overline{\nu\mu}}
-\nabla_{\overline\mu}S_{\overline{\nu ij}} +(\alpha-\beta)(\partial_{\overline\nu}\rho S_{\overline{\mu ij}}
+\partial_{\overline\mu}\rho S_{\overline{\nu ij}})\nonumber\\
&&
-\alpha\partial^{\overline\lambda}\rho S_{\overline{\lambda ij}}\eta_{\overline{\mu\nu}}
- (S_{\overline \mu}S_{\overline\nu})_{\overline {ij}}
-[S_{\overline\nu},Q_{\overline\mu}]_{\overline{ij}}
\Big)\nonumber\\
\hat R_{\overline{\lambda \mu i\nu }}&=&
\frac{1}{2}e^{(\beta-3\alpha)\rho} \Big(
e_{\overline i j}\nabla_{\overline \nu}F^j_{\overline{\lambda\mu}}+
(\beta-\alpha)(
2\partial_{\overline\nu}\rho F_{\overline{i\lambda\mu}} +\partial_{\overline\mu}\rho F_{\overline{i\lambda\nu}}
+\partial_{\overline\lambda}\rho F_{\overline{i\nu\mu}})  \nonumber\\
&&
+\alpha \partial^{\overline \rho}\rho( F_{\overline{i\lambda\rho}}\eta_{\overline{\mu\nu}}
- F_{\overline{i\mu\rho}}\eta_{\overline{\lambda\nu}})
+2S_{\overline{\nu ij}} F^{\overline j}_{\overline{\lambda\mu}}
+S_{\overline{\mu ij}} F^{\overline j}_{\overline{\lambda\nu}}
+S_{\overline{\lambda ij}} F^{\overline j}_{\overline{\mu\nu}}
\Big)\nonumber\\
\hat R_{\overline{\lambda\rho\mu\nu}}&=&e^{-2\alpha\rho}R_{\overline{\lambda\rho\mu\nu}}
-\frac{1}{2}e^{2(\beta-2\alpha)\rho}F^{\overline i}_{\overline{\lambda\rho}}
F^{\overline i}_{\overline{\mu\nu}} - \frac{1}{4}e^{2(\beta-2\alpha)\rho}
(F^{\overline i}_{\overline{\lambda\mu}} F^{\overline i}_{\overline{\rho\nu}}
-F^{\overline i}_{\overline{\lambda\nu}} F^{\overline i}_{\overline{\rho\mu}})\nonumber\\
&&
-\alpha e^{-2\alpha\rho}\big(
\nabla_{\overline\lambda}\partial_{\overline\mu}\rho\eta_{\overline{\rho\nu}}
-\nabla_{\overline\lambda}\partial_{\overline\nu}\rho\eta_{\overline{\rho\mu}}
-\nabla_{\overline\rho}\partial_{\overline\mu}\rho\eta_{\overline{\lambda\nu}}
+\nabla_{\overline\rho}\partial_{\overline\nu}\rho\eta_{\overline{\lambda\mu}}\big)\nonumber\\
&&
+\alpha^2 e^{-2\alpha\rho}\big(
\partial_{\overline\lambda}\rho\partial_{\overline\mu}\rho\eta_{\overline{\rho\nu}}
-\partial_{\overline\lambda}\rho\partial_{\overline\nu}\rho\eta_{\overline{\rho\mu}}
+\partial_{\overline\rho}\rho\partial_{\overline\nu}\rho\eta_{\overline{\lambda\mu}}
-\partial_{\overline\rho}\rho\partial_{\overline\mu}\rho\eta_{\overline{\lambda\nu}}
\nonumber\\
&&+(\partial\rho)^2(\eta_{\overline{\lambda\nu}}\eta_{\overline{\rho\mu}}-\eta_{\overline{\lambda\mu}}\eta_{\overline{\rho\nu}})
\big)
\label{Rcomp}
\end{eqnarray}

From these expressions we find
\begin{eqnarray}
\hat R_{\overline {ij}}&=&e^{-2\alpha\rho}\Big(-\beta\nabla^2\rho\delta_{\overline{ij}}
-((d-2)\alpha\beta+n\beta^2)(\partial\rho)^2\delta_{\overline{ij}}
-\frac{1}{4}e^{2(\beta-\alpha)\rho}{\rm Tr}(F_{\overline i}F_{\overline j})\nonumber\\
&&
-\nabla^{\overline \mu}S_{\overline{\mu ij}}
- ((d-2)\alpha+n\beta)\partial^{\overline\mu}\rho S_{\overline{\mu ij}}
\Big)\nonumber\\
\hat R_{\overline {\mu i}}&=&\frac{1}{2}e^{(\beta-3\alpha)\rho}
\Big(e_{\overline i j}\nabla^{\overline\nu}F^j_{\overline{\mu\nu}}+2S^{\overline \nu}_{\overline ij}F^{\overline j}_{\overline{\mu\nu}}+((d-4)\alpha-\beta(n-4))\partial^{\overline\nu}\rho F_{\overline{i\mu\nu}}
\Big)\nonumber\\
\hat R_{\overline {\mu\nu} }&=&e^{-2\alpha\rho}
\Big(R_{\overline{\mu\nu}}+\frac{1}{2}e^{2(\beta-\alpha)\rho}(F^{\overline i}F_{\overline i})_{\overline {\mu\nu}}
-(n\beta+(d-2)\alpha)\nabla_{\overline\mu}\partial_{\overline\nu}\rho\nonumber\\
&&
+((d-2)\alpha^2+(2\alpha\beta-\beta^2)n)\partial_{\overline \mu}\rho\partial_{\overline\nu}\rho
-((d-2)\alpha^2+\alpha\beta n)(\partial\rho)^2\eta_{\overline{\mu\nu}}\nonumber\\
&&-\alpha\nabla^2\rho\eta_{\overline{\mu\nu}}-{\rm Tr}(S_{\overline \mu}S_{\overline\nu})\Big)\nonumber\\
\end{eqnarray}
and
\begin{eqnarray}
\hat R &=& e^{-2\alpha\rho}\Big(R - \frac{1}{4}e^{2(\beta-\alpha)\rho}
 G_{ij}F^i_{{\mu\nu}} F^{{j\mu\nu}} - S_{\mu i}^{\ \ j}S^{\mu i}_j-\gamma^2(\partial\rho)^2\nonumber\\
&&
-2(n\beta+(d-1)\alpha)\nabla^2\rho
\Big)
\end{eqnarray}
where
\begin{equation}
\gamma^2 = (d-1)(d-2)\alpha^2+(2dn-4n)\alpha\beta+n(n+1)\beta^2
\end{equation}
To reduce to Einstein
frame we require that
\begin{equation}
\beta  = -\Big({d-2\over D-d}\Big)\alpha
\end{equation}
Finally we fix $\alpha$ by taking the standard normalization for the kinetic
energy of a scalar field, $\gamma^2=1/2$. In this paper we are interested
in taking $d=3$ in which case these formulae simplify to
\begin{equation}
\alpha^2 = \frac{1}{2}\frac{n}{n+1}\ ,\qquad \beta = -\frac{1}{n}\alpha
\end{equation}

\section*{Appendix B}

Let review some elementary facts about $G/H$ cosets.
We assume  that $G$ is a finite dimensional semi-simple Lie algebra. It consists of the elements
 $\underline H$, which form the Cartan subalgebra, and
$E_{\underline \alpha}$ subject to the relations
\begin{equation}
[\underline H,E_{\underline \alpha}] = {\underline \alpha}E_{\underline \alpha}
\end{equation}
In this appendix we use an underline to denote quantities such as roots vectors, however,
in the body of the paper such quantities have underlines only if they
belong to manifest
$SL(n)$ symmetry associated with the torus reduction and  arrows if
they belong to the group  associated to the enhanced   coset symmetry.
The roots $\underline \alpha$ can be split into positive and negative roots,
which we denote by $\underline\alpha >0$ and $\underline\alpha <0$ respectively.

Such  algebras admit the Cartan  involution
\begin{equation}
\tau:(E_{\underline \alpha},E_{-\underline \alpha},\underline H)\to -(E_{-\underline \alpha},E_{\underline \alpha},\underline H)
\end{equation}
where $\underline \alpha>0$. This is a group automorphism and so
obeys $\tau(g_1g_2) =\tau(g_1)\tau(g_2)$ for any two group elements
$g_1$ and $g_2$. This allows us to construct a generalized transpose
$A^\#=-\tau(A)$. We then take $H$ to be the sub-algebra invariant
under the Cartan involution. It terms of  the $\#$ operation it
consist of the elements of the algebra for which $A^\#=-A$,  or in
terms of group elements  those that obey $h^\#=h^{-1}$.

We consider elements of the group which depend on space-time to transform as
\begin{equation}
g\to g_0 g h
\label{cosettrans}
\end{equation}
where $g_0\in G$ is a rigid, that is  constant group element and $h\in H$ is space-time dependent.
Using this latter local invariance one can take an element $g\in G$ to be
of the form
\begin{equation}
g = e^{\sum_{\underline \alpha>0}\chi_{\underline\alpha}E_{\underline \alpha}}
\ e^{-\frac{1}{\sqrt{2}}\underline\phi \cdot \underline H}
\end{equation}

To construct the dynamics it is useful to use  the Cartan forms
\begin{equation}
{\cal V}_\mu=g^{-1}\partial_\mu g = -\frac{1}{\sqrt{2}}\partial_\mu \underline\phi \cdot \underline H
+\sum_{\underline \alpha>0} e^{\frac{1}{\sqrt{2}}\underline\phi\cdot \underline\alpha}\partial_\mu\chi_{\underline\alpha}E_{\underline\alpha}
\end{equation}
up to higher order terms in $\chi_{\underline\alpha}$. Under the
transformation (\ref{cosettrans})  the Cartan forms transform as
${\cal V}_\mu  \to h^{-1}{\cal V}_\mu  h+h^{-1}{\partial}_\mu  h$.
Using the Cartan involution we can construct the objects
\begin{equation}
{\cal S}_\mu=\frac{1}{2}{\cal V}_\mu+\frac{1}{2}\tau ({\cal V}_\mu), \qquad \omega_\mu={1\over
2}({\cal V}_\mu-\tau({\cal V}_\mu))
\end{equation}
which transform as
\begin{equation}
{\cal S}_\mu\to h^{-1}{\cal S} _\mu h\ ,\qquad \omega_\mu \to   h^{-1}\omega_\mu h+  h^{-1}\partial_\mu h
\end{equation}
Under the $\#$ operation we find that
$\omega_\mu^\#=\omega_\mu$ and ${\cal S}_\mu^\#=-{\cal S}_\mu$ and
hence they lie in the subalgebra $H$ and its complement $G-H$ respectively.
The first such quantity is given by
\begin{equation}
{\cal  S}_\mu = -\frac{1}{\sqrt{2}}\partial_\mu \underline\phi \cdot \underline H
+\frac{1}{2}\sum_{\underline \alpha>0} e^{\frac{1}{\sqrt{2}}\underline\phi\cdot \underline\alpha}\partial_\mu\chi_{\underline\alpha}(E_{\underline\alpha}+E_{\underline\alpha}^\#)
\end{equation}
We can then construct an invariant Lagrangian by taking
\begin{eqnarray}
L &=& -{\rm Tr}({\cal S}_\mu {\cal S}^\mu)\nonumber\\
&=& -\frac{1}{2}\partial_\mu \underline\phi \cdot\partial^\mu \underline\phi
-\frac{1}{2}\sum_{\underline \alpha>0} e^{\sqrt{2}\underline\phi\cdot \underline\alpha}
\partial_\mu\chi_{\underline\alpha}\partial^\mu\chi_{\underline\alpha}
\label{cosetact}
\end{eqnarray}
up to higher order terms in $\chi_{\underline\alpha}$.

There is an alternative way to construct the same Lagrangian. This
time one starts with
\begin{eqnarray}
{\cal M} &=& gg^\#\nonumber\\
&=&  e^{\sum_{\underline \alpha>0}\chi_{\underline\alpha}E_{\underline \alpha}}
\ e^{-\sqrt{2}\underline\phi \cdot \underline H}
\ e^{\sum_{\underline \alpha> 0}\chi_{\underline\alpha}E_{-\underline \alpha}}
\nonumber\\
{\cal M}^{-1} &=& e^{-\sum_{\underline \alpha>0}
\chi_{\underline\alpha}E_{-\underline \alpha}}
\ e^{\sqrt{2}\underline\phi \cdot \underline H}
\ e^{-\sum_{\underline \alpha> 0}\chi_{\underline\alpha}E_{\underline \alpha}}\nonumber\\
\label{G}
\end{eqnarray}
Under $g\to g_0 g h$ we see that ${\cal M} \to g_0{\cal M} g_0^\#$ since the action of
$\#$ in the Lie algebra lifts to  $h^\# =h^{-1}$ for $h\in H$.
Thus another possible invariant Lagrangian is
\begin{equation}
L = -\frac{1}{4}{\rm Tr}(\partial_\mu {\cal M} \partial_\mu {\cal M}^{-1})
\end{equation}
but in fact  this is the same as  (\ref{cosetact}).

When $G=SL(n)$, $\#$ is simply the transpose so that
$H=SO(n)$ and ${\cal S}_\mu$ is  the symmetric
part of  $g^{-1}\partial_\mu g$.

\section*{Appendix C}

In this appendix we will give an account of certain aspects  of the
theory of group representations that are required in this paper. We
recall that a representation $R$ of a group $G$ consists of a vector  space
$V$ and a set of operators $U(g),\ \forall g\in G$ which act on $V$,
namely
$|\psi>
\to U(g)|\psi>$ such that
$ U(g_1g_2) =U(g_1)U(g_2)$.
We will take the algebra $G$ to be a finite dimensional semi-simple  and
simply laced. The states in the representation can be chosen so as  to be
eigenstates of
$\underline H$. The eigenvalues being called the weights. It can be  shown
that  the weights of $G$ belong to  the dual lattice to
the lattice of roots, {\it i.e.} a weight $\underline w$ satisfies
\begin{equation}
{\underline w}\cdot{\underline \alpha_a}\in {\bf Z}
\end{equation}
for the simple roots  $\underline \alpha_a$. The representations of
interest to us are finite dimensional and so must have a highest weight
$\underline \lambda$ which is the one such that $\underline
\lambda+\underline \alpha_a$ is not a weight for all simple roots
$\underline \alpha_a$. The representations will  also have a lowest root
denoted $\underline \mu$. Of particular interest are the fundamental
representations which are those whose highest weights
$\underline \lambda^a$ obey the relation
\begin{equation}
\underline\lambda^a\cdot\underline\alpha_b = \delta^a_{b}
\end{equation}
for all simple roots  $\underline \alpha_a$.
The roots are themselves weights and these correspond to
the adjoint representation,  whose highest weight we will
denote by $\underline \theta$.

For $SL(n)$, {\it i.e.} $A_{n-1}$,
the fundamental weights $\underline \lambda^a$ satisfy
\begin{equation}
\underline\lambda^a\cdot\underline \lambda^b = a(n-b)/n
\end{equation}
for $b \ge a$.
The representation with highest weight $\underline \lambda^{n-k}$
is realized on a tensor with $k$ totally anti-symmetrized
superscript indices, i.e. $T^{i_1\ldots i_k}=T^{[i_1\ldots i_k]}$.
Using the group invariant epsilon symbol $\epsilon ^{i_1\ldots i_n}$,  this
representation is equivalent to taking a tensor with
$n-k$ lowered indices.

Given any simple root one may carry out its Weyl reflection on any
weight
\begin{equation}
S_\alpha(w)=\underline w-(\underline \alpha\cdot \underline w)\underline
\alpha
\end{equation}
The collection of all such reflections is called the Weyl group and
it can be shown that any member of it can be written in terms of a
product of Weyl reflections in the simple roots. Although the
precise decomposition of a given element of the Weyl group is not
unique its  length is defined to be the smallest number of simple
root reflections required. However,   there does exist a unique Weyl
reflection, denoted $W_0$,  that has the longest length. This
element  obeys $W_0^2=1$, takes the positive simple roots to
negative simple roots and its length is the same as the number of
positive roots.  As a result, $-W_0$ exchanges the positive simple
roots with each other and,  as  Weyl transformations preserve the
scalar product, it must also preserve  the Cartan matrix.
Consequently,  it must lead to  an automorphism of the Dynkin
diagram. Given any representation of $G$ the highest and lowest
weights are related by
\begin{equation} \underline \mu
=W_0\underline \lambda
\end{equation} Given the definition of the
fundamental weights and carrying out a Weyl transformation
$W_0$, we may conclude that the negative of the highest and lowest
weights of a given fundamental representation are the lowest  and
highest representation  of one of the other fundamental
representations. It is always the case that the two
representations have the same dimension. However it can happen that
a fundamental representation is self-dual.

For $SL(n)$ $W_0=(S_{\underline\alpha_1}\ldots S_{\underline\alpha_
{n-1}})(S_{\underline\alpha_1}\ldots S_{\underline\alpha_{n-2}})
\ldots (S_{\underline\alpha_1}S_{\underline\alpha_2})S_{\underline
\alpha_1}$ and  one finds that, in this case,
\begin{equation}
W_0 \underline \lambda _{n-k}=  \underline \mu _{n-k}=- \underline
\lambda _{k} \iff W_0 \underline \mu _{n-k}=  \underline \lambda
_{n-k}=- \underline \mu _{k} .
\end{equation}
This result also follows from the above remarks on $W_0$ as it must
take a fundamental representation to a fundamental representation
and correspond to an automorphism of the Dynkin diagram which in
this case is just takes the nodes $k$ to $n-k$.

Given a representation acting on $|\chi> \in V$ we may  consider
the dual representation $R_D$  that is carried by the space of linear
functionals, denoted  $V^*$,  acting on $V$. The group action is defined
by
\begin{equation}
<\chi_D | \to <\chi_D |U(g^{-1}) ,\ \forall g \in G,\ <\chi_D |  \in
V^*
\label{thisone}
\end{equation}
We note that $<\chi_D |\chi>$ is $G$-invariant. Since the linear
functionals  carry a representation we may also choose a basis them
that is  labeled by the weights. It is easy to see that  a linear
functional with a weight $\underline w$ only has a non-zero result
on a state with weight $-\underline w$. A little further thought
allows one to conclude that  if the representation $R $ has highest
and lowest weight $\underline \lambda$ and $\underline \mu$
respectively then the dual representation has a highest weight
$-\underline \mu$ and lowest weight $-\underline \lambda$. Indeed
the dual representation has the same dimension as the original
representation. For the case of $SL(n)$, {\it i.e.} $A_{n-1}$, if
the representation $R$ is the fundamental representation with
highest weight $\underline \lambda^k$ then it follows from equation
(\ref{thisone}) that the dual representation is the fundamental
representation with highest weight $\underline \lambda^{n-k}$. Thus
the representations carried by $T^{i_1\ldots i_{(n-k)}}$ is dual to
the representation carried by $T^{i_1\ldots i_k} $ or in that latter
case carried by $T_{i_1\ldots i_{(n-k)}}$ as might be expected.

Given the representation $R$ and any automorphism of the group  $\tau$
({\it i.e.} $\tau(g_1g_2)=\tau (g_1) \tau(g_2)$) we
may also define a twisted representation $R_T$ on the same vector space
$V$ by
\begin{equation}
|\phi_T> \to U(\tau(g))|\phi_T>\ \forall g\  \in G,\ |\phi_T>\ \in \ V
\end{equation}
The case of interest to us in this paper  is when we take the
automorphism to be the Cartan involution  which we also denoted by
$\tau$. It is easy to see that if the representation $R$  has
highest and lowest  weight $\underline \lambda$ and $\underline \mu$
respectively then the dual representation has a highest weight
$-\underline \mu$ and lowest weight $-\underline \lambda$ and so the
twisted representation is isomorphic to  the dual representation.

In the first systematic account of the theory of non-linear realizations
\cite{Callan:1969sn} it was explained that one can convert a linear realization into a
non-linear realization. In particular, given a non-linear realization
with group element $g$ which transforms as $g\to g_0 g h$ and any linear
representation
$R$ we find that
\begin{equation}
U(g^{-1})|\psi>\to U(h^{-1}) U(g^{-1})|\psi>
\end{equation}
We now want to generalize this construction to the situation we will
encounter in this paper. Given the dual representation $R_D$
(\ref{thisone}) we find that
\begin{equation}
<\chi_D | U(g) \to <\chi_D | U(g) U(h)
\label{app}
\end{equation}
and if we take the Cartan twisted representation of equation
(\ref{thisone}) then
\begin{equation}
U(g^{\#} )|\phi_T> \to U(h^{-1}) U(g^{\#}) |\phi_T>
\end{equation}
It follows that if we consider any representation $|\chi>$, then
$<\chi_D | U(g g^{\#})|\chi_T>$ is inert under local and rigid
transformations. In this latter expression $|\chi_T>$ and $<\chi_D
|$ are the Cartan twisted and dual representations derived from
$|\chi>$ respectively.  We note that the twisted and dual
representations are actually isomorphic to each other. As we have
seen,  this quantity  plays a central role in this paper. Similarly
one can show that
\begin{equation}
<(\chi_T)_D|U({\cal M}^{-1})|\chi> \label{overhere}
\end{equation}
where ${\cal M}=g g^{\#}$, is invariant under local and rigid
transformations.

We note that the two representations appearing in equation (\ref{overhere})  are
isomorphic. If we take
as a basis of this representation  to be given by the states
$|I,\underline\lambda>$  we may write $|\chi>$ as $|\chi>=\sum_I
\chi_I|I,\underline\lambda>$, and similarly for  $<(\chi_T)_D|$ then the above
expression becomes
\begin{equation}
<(\chi_T)_D|U({\cal M}^{-1})|\chi>=\sum _{IJ} \chi_I <I,\underline\lambda |
U(M^{-1}) |J,\underline\lambda>\chi_J
\end{equation}

Indeed, if we consider the group to be $SL(n)$ and take $|\chi>$ to be the
representation with highest weight $\underline\lambda^1$, that is in the
vector representation,   then
$<(\chi_T)_D|$ is also an $SL(n)$ vector. As a result,  we find that
\begin{equation}
<(\chi_T)_D|U({\cal M}^{-1}|\chi>=\sum _{IJ} \chi_I g^{IJ} \chi_J
\end{equation}
as in this case $g^{IJ}=<I,\underline\lambda^1 | U(M^{-1})
|J,\underline\lambda^1>$ is the metric on the torus as we have shown in section 2.

\end{document}